\documentclass[aps, twocolumn,
superscriptaddress,showpacs,showkeys,amsmath,amssymb,floatfix]{revtex4}
\usepackage{color}
\usepackage{graphicx}
\usepackage{epsfig}
\usepackage{dcolumn}
\usepackage{bm}
\usepackage{mathtools}
\usepackage{amssymb}
\usepackage{bmpsize}
\usepackage{calrsfs}
\DeclareMathAlphabet{\pazocal}{OMS}{zplm}{m}{n} 
\usepackage{amsmath}
\usepackage{subfigure}
\newcommand{\be}{\begin{equation}}
\newcommand{\ee}{\end{equation}}
\newcommand{\bea}{\begin{eqnarray}}
\newcommand{\eea}{\end{eqnarray}}

\newcommand{\tK}{{\tilde K}}

\newcommand{\tF}{{\tilde F}}
\newcommand{\tf}{{\tilde f}}

\newcommand{\tg}{{\tilde g}}

\newcommand{\tM}{{\tilde M}}

\newcommand{\tk}{{\tilde k}}
\newcommand{\tw}{{\tilde w}}

\newcommand{\pro}{\partial}

\newcommand{\bfA}{{\mathbf A}}

\newcommand{\bfr}{{\mathbf r}}

\newcommand{\bfw}{{\mathbf w}}

\newcommand{\ba}{\begin{array}}
\newcommand{\ea}{\end{array}}

\newcommand{\nn}{\nonumber}

\newcommand{\Ka}{\mathcal{K}}

\newcommand{\Da}{\mathcal{D}}

\newcommand{\Ab}{\pazocal{A}}
\newcommand{\mcF}{\mathcal{F}}

\begin{document} 
\title{Color confinement and color singlet structure of quantum states in Yang-Mills theory}
\author{D.G. Pak}
\affiliation{Institute of Theoretical Physics, Chinese Academy of Sciences, Beijing 100190, China}
\affiliation{School of Education, Bukkyo University, Kyoto 603-8301, Japan}
\affiliation{Chern Institute of Mathematics, Nankai University, Tianjin 300071, China} 
\author{Rong-Gen Cai}
\affiliation{Institute of Theoretical Physics, Chinese Academy of Sciences, Beijing 100190, China}
\author{Takuya Tsukioka}
\affiliation{School of Education, Bukkyo University, Kyoto 603-8301, Japan}
\author{Pengming Zhang}   
\affiliation{Institute of Modern Physics, Chinese Academy of Sciences,
Lanzhou 730000, China}
\affiliation{School of Physics and Astronomy,
 Sun Yat-sen University, Zhuhai, 519000, China}
\author{Yu-Feng Zhou}   
\affiliation{Institute of Theoretical Physics, Chinese Academy of Sciences, Beijing 100190, China}
\begin{abstract}
We consider two fundamental long-standing problems in quantum chromodynamics (QCD):
the origin of color confinement and structure of a true vacuum and color singlet quantum states.
There is a common belief that resolution to these problems needs a knowledge of a strict non-perturbative 
quantum Yang-Mills theory and new ideas.
Our principal idea in resolving these problems is that structure of color confinement and color singlet quantum states 
must be determined by a Weyl symmetry which is an intrinsic symmetry of the Yang-Mills gauge theory. 
Following this idea we construct for the first time a space of color singlet one particle quantum states for 
primary gluons and quarks, and reveal the structure of color confinement in quantum Yang-Mills theory. 
As an application we demonstrate formation of physical observables in a pure QCD, pure glueballs. 
\end{abstract}
\maketitle
     
\section{Introduction}     

Color confinement represents the most amazing phenomenon in quantum chromodynamics.
Despite on tremendous progress in QCD since its invention the origin and mechanism of color confinement 
remains unclear. The first deep insight on the nature of the color confinement was revealed long time ago
in \cite{thooft81, polyakov77} where it was stressed that the color confinement phenomenon
is intimately related to the gauge invariance of the vacuum. Based on  this `t Hooft conjectured that QCD 
vacuum structure in the confinement phase must be described by Abelian fields which supposed to be color neutral.
First scenarios for the color confinement mechanism based on monopole condensation 
were proposed in \cite{nambu74, mandelstam76} and developed in subsequent studies. 
Existence of a stable QCD vacuum is another long-standing problem 
since finding in 1977 by Savvidy \cite{savv} that a non-trivial QCD vacuum can be generated by radiative corrections. 
Quantum instability of Savvidy vacuum established in the seminal paper by Nielsen and Olesen \cite{N-O} triggers 
active extensive searches of stable vacuum field configurations in the subsequent several decades.
An explicit microscopic structure of stable QCD vacuum has been remained unknown despite on significant progress in
approximate description of the vacuum in various approaches 
\cite{N-O,niel-oles2,amb-oles2,chernodub14,centervort1,centervort2,centervort3,diak-petrov}.

In the present paper we propose a novel approach to resolution of the problems of color confinement 
and vacuum stability elaborating an idea that  Weyl symmetry of the color group $SU(3)$ is a principal symmetry 
which determines all color attributes  of vacuum and quantum states, and provides microscopic description of a 
true stable color invariant vacuum.
In  Section II we formulate main requirements that solutions must meet for proper definition of one particle quantum states.
Based on this we construct an ansatz for Weyl symmetric stationary solutions of magnetic and dual electric type  
and construct solutions to equations of motion of $SU(3)$ Yang-Mils theory. We demonstrate that obtained 
Weyl symmetric solutions represent fixed points in the configuration field space under Weyl transformation and possess a
vanishing total color charge. We prove that each solution space with a fixed set of quantum numbers is one-dimensional
and provides a color singlet one particle quantum state. The Weyl symmetric solutions manifest an Abelian dominance effect,
which allows to classify non-Abelian solutions and construct a full Hilbert space of color singlet 
quantum states, at least in principle. In Section III we apply the Weyl symmetric ansatz to Dirac equation in 
QCD with one flavor quark. Obtained results reveal unexpected feature: there are three independent 
Weyl symmetric quark solutions: two color quarks and one free colorless quark, contrary to conventional wisdom that one has 
three color quarks. 
 Section IV is devoted to a persistent problem in all known QCD vacuum models, the quantum vacuum stability against 
 quantum fluctuations.
 We prove that vacuum Abelian Weyl symmetric solutions are stable under quantum gluon fluctuations. 
 Thus, the Weyl symmetric solutions
 provide microscopic description of stable vacuum gluon and quark condensates.
 In Section V we consider equations of motion corresponding to quantum one-loop effective action of a pure QCD. 
 We show that  equations with quantum corrections admit a stationary solution localized in a finite space region.
It has been demonstrated that the lightest pure glueball is formed due to interaction of the primary gluon with a
corresponding generated vacuum gluon condensate.
A qualitative spectrum of lightest scalar glueballs is calculated  in agreement with the Regge theory. In Appendix 
we consider a reduced system of equations for propagating off-diagonal and Abelian modes $K_{2,4}$ in $1+1$
dimensional space-time. We demonstrate that energy conservation law in a finite space domain leads to
correlation of solution modes $K_{2,4}$. As a result the Weyl symmetric solution possesses only one independent dynamic
degree of freedom and leads to a color singlet quantum state after standard canonical  quantization.
Possible implications of our results are enclosed in the last section.

\section{Weyl symmetric solutions}
\subsection{A basic idea and requirements to classical soluitons}

Yang-Mills theory is formulated on a basis of a strict mathematical scheme of fiber bundle supplied with a 
structural group. We consider Yang-Mills theory with a gauge group $SU(3)$ which represents a classical
theory for a conventional quantum chromodynamics.
The structure of the gauge group is the only mathematical structure which determines all symmetry properties
of the theory defined by a standard Yang-Mills Lagrangian. 
An important role of the gauge principle is that gauge symmetry defines dynamics of the physical system by means 
of Euler equations of motion, and governs all properties of corresponding solutions which describe the classical system.
From a formal point of view the gauge symmetry looks redundant and usually during quantization procedure
one has to  fix the gauge symmetry to select one field representative in each gauge equivalence class of fields.
This can be done by numerous ways, and a basis set of dynamical solutions can be chosen arbitrarily.
A consistent quantization procedure provides physical quantities  to be independent on a choice of gauge fixing condition
for quantum fields. This is true for quantum virtual fields since the functional integration is performed over space of all 
possible quantum fluctuations in a gauge invariant manner, and it is clear, that choice of basis fields is not important.
However, in general, the physical properties of a system depends on which class of 
classical solutions is selected contrary a commonly accepted opinion that all gauges are equivalent. 
Especially this is important in non-Abelian theory which does not admit linear superposition principle, and typically 
it is not possible to construct a complete basis in the space of non-linear solutions. So that an improper choice of a basis 
of classical solutions for construction of a Hilbert space of quantum states will lead to inconsistent concepts of particles, 
 and physical observables. 
 
We adopt the following requirements to classical solutions in the Yang-Mills theory which
are used for further description and construction of vacuum and quantum states:\\
(i) after fixing the local gauge symmetry only symmetries corresponding to global or finite subgroups of the original 
structural group survive. We follow an idea that a Weyl group of $SU(3)$ is the only proper color symmetry group
which survives after gauge fixing, and determine all color attributes of solutions.
We require that solutions describing vacuum and quantum states
 must be defined by an ansatz invariant under Weyl transformation. Global color symmetries are not acceptable 
 since they imply spontaneous color symmetry breaking which prevents appearance of color confinement phase. \\
(ii) Consistence with quantum mechanical principles implies that at microscopic space-time level the solutions must 
depend on time and admit stationary classical states with a conserved energy. This was observed in early papers
 \cite{niel-oles2, amb-oles2}, so the static solutions must be excluded unless the gauge symmetry is broken.\\
(iii) solutions must admit localization of particle states in a finite space regions, since all hadrons are localized objects and 
there is no massless hadrons like  free photons which are not localized in space.\\
(iv) solutions must be exact solutions to exact equations of motion, otherwise some important non-perturbative
 features will be lost.\\
(v) vacuum solutions describing the microscopic structure of vacuum gluon and quark condensate must be stable against
quantum fluctuations and possess classical stability, saddle-point solutions are not acceptable.\\
(vi) solutions must be regular, possess finite energy density and a conserved total energy inside hadrons.\\
Based on these requirements we construct Weyl symmetric solutions which provide microscopic description of the vacuum and
color singlet structure of the Hilbert space of quantum states.
 
\subsection{Weyl symmetric $SU(3)$ ansatz \\
for stationary magnetic type solutions}

 Let us consider first the $SU(2)$ Yang-Mills theory with a standard Lagrangian  
    $(\mu,\nu=0,1,2,3; a=1,2,3)$
   \bea
&&{\cal L}_{0} = -\dfrac{1}{4} F_{\mu\nu}^a F^{a\mu\nu}. \label{Lagr0}
\eea
   A generalized Dashen-Hasslacher-Neveu (DHN)
   ansatz \cite{DHN} for time dependent axially symmetric solutions of magnetic type
   is defined by means of the following non-vanishing components of the gauge potential $A_\mu^a$ 
   \cite{plb2018}
   \begin{align}
A_t^2&=K_0(r,\theta,t),& A_r^2&= K_1(r,\theta,t),~~A_\theta^2= K_2(r,\theta,t), \nn \\
A_\varphi^3&=K_3(r,\theta,t), & A_\varphi^1&= K_4(r,\theta,t).  \label{genDHN}
\end{align}
The ansatz is invariant under residual $U(1)$ transformations with a gauge
   parameter  $\lambda(r,\theta,t)$ \cite{Manton78,RR,KKB} 
\begin{align}
K'_0&=K_0+\pro_t \lambda,~~K'_1=K_1+\pro_r \lambda,~~K'_2=K_2+\pro_\theta \lambda, \nn \\
K_3'&=K_3 \cos  \lambda+K_4\sin  \lambda, \label{residual} \\
K_4'&=K_4 \cos  \lambda-K_3 \sin  \lambda. \nn
\end{align}
One can fix the local $U(1)$ symmetry by adding a gauge fixing term ${\cal L}_{\rm gf}$ to the original
Yang-Mills Lagrangian
 \be
{\cal L}_{\rm gf}=-\dfrac{1}{2} (\pro_t K_0-\pro_r K_1-\dfrac{1}{r^2} \pro_\theta K_2)^2. \label{gfterm}
\ee
After fixing a gauge the Yang-Mills Lagrangian is still invariant under global color $SO(2)$ transformations
with a constant parameter $\lambda$ in (\ref{residual}). One can fix the global symmetry and define a minimal ansatz
by imposing a constraint 
$K_3=c_3 K_4$ (we set $c_3=\sqrt 2/2$ without loss of generality).
With this, five equations of motion for the fields $K_{\hat i}$ ($\hat i=0,1,2,4$)reduce
to four equations 
\begin{align}
r^2 \pro_t^2 K_1-r^2 \pro_r^2 K_1-\pro_\theta^2 K_1+2r(\pro_t K_0-\pro_r K_1)& \nn \\
+\cot\theta(\pro_r K_2-\pro_\theta K_1)+\dfrac{9}{2}\csc^2\theta K_4^2 K_1&=0,\label{eq1} \\
r^2\pro_t^2 K_2-r^2 \pro_r^2 K_2-\pro_\theta^2 K_2+r^2 \cot\theta(\pro_t K_0-\pro_r K_1)&\nn \\
-\cot\theta \pro_\theta K_2+\dfrac{9}{2}\csc^2\theta K_4^2 K_2&=0,  \label{eq2} \\
r^2 \pro_t^2 K_4-r^2 \pro_r^2 K_4-\pro_\theta^2 K_4+\cot\theta\pro_\theta K_4&\nn \\
+3r^2(K_1^2-K_0^2)K_4+3K_2^2K_4&=0,\label{eq3} \\
r^2 \pro_t^2 K_0-r^2 \pro_r^2 K_0-\pro_\theta^2 K_0+2r(\pro_t K_1-\pro_r  K_0)&\nn \\
+\cot\theta (\pro_t K_2-\pro_\theta K_0)+ \dfrac{9}{2}\csc^2\theta K_4^2 K_0&=0, \label{eq4}
\end{align}
and one quadratic constraint
\begin{align}
2 r^2 (K_0\pro_t K_4-K_1 \pro_r K_4 )+ K_2 (\cot \theta K_4-2 \pro_\theta K_4)&\nn \\
+K_4 (-\pro_\theta K_2 +r^2 (\pro_t K_0-\pro_r K_1))&=0.  \label{constr1}
\end{align}
A total Yang-Mills Lagrangian with gauge fixing terms is simplified as follows
\begin{align}
{\cal L}_{\rm tot}&={\cal L}_0(K)+{\cal L}_{\rm gf}\nn \\
&=\dfrac{1}{2r^2}\Big [r^2(\pro_t K_1 -\pro_r K_0)^2-(\pro_\theta K_1)^2+(\pro_\theta K_0)^2\Big ]\nn\\
&+\dfrac{1}{2 r^2} \Big [ \pro_t K_2 (\pro_t K_2- \pro_\theta K_0) - \pro_r K_2(\pro_r K_2-\pro_\theta K_1) \Big ]\nn \\
&+\dfrac{3}{4r^4 \sin^2 \theta}\Big [ r^2 ((\pro_t K_4)^2-(\pro_r K_4)^2) -(\pro_\theta K_4)^2 \Big] \nn \\
&      -\dfrac{3}{4 r^4\sin^2 \theta} \Big [K_4^2 (K_2^2+r^2 (K_1^2-K_0^2)) \Big ]. \label{Lred}
\end{align}
Using the $SU(2)$ ansatz (\ref{genDHN}) we construct a Weyl symmetric ansatz for 
$SU(3)$ Yang-Mills theory by setting non-vanishing components of the gauge potential $A_\mu^a$
corresponding to $I,U,V$ type subgroups $SU(2)$ as follows \footnote{A Weyl symmetric ansatz
in the present paper is different from the ansatz in \cite{plb2018}.
A modified ansatz (\ref{SU3DHN}, \ref{reduction1}) implies that fields $K_{\hat i}$ represent fixed points under 
Weyl transformation, this provides color singlet structure of quantum states. Besides, all cubic interaction terms
 in the Lagrangian are mutually canceled irrespective of the
relationship between Abelian fields $K_3$ and $K_4$.}
\begin{align} 
I:~~A_t^2&=K_0, &A_r^2&=K_1, & A_\theta^2&=K_2, & A_\varphi^1&=K_4,\nn \\
U:~A_t^5&=-Q_0, & A_r^5&=-Q_1, & A_\theta^5&=-Q_2, & A_\varphi^4&=Q_4, \nn \\
V:~A_t^7&=S_0, & A_r^7&=S_1, & A_\theta^7&=S_2, & A_\varphi^6&=S_4,\nn \\
{\Ab}_\varphi^p&=A_\varphi^\alpha r_\alpha^{\,p}, & 
A_\varphi^3&=K_3, & A_\varphi^8&= K_8, \label{SU3DHN} 
\end{align}
where $r_\alpha^{\,p}$ $(p=I,U,V,~ \alpha=3,8)$ are root vectors
$\bfr^{1}=(1,0)$, $\bfr^{2}=(-1/2,\sqrt 3/2)$, $\bfr^{3}=(-1/2,-\sqrt 3/2)$.
The Weyl group acts on color components $A_\mu^a$ as a symmetric permutation group $S_3$ and realizes 
eight dimensional reducible representation.
One can define a minimal ansatz by imposing additional constraints ($i=0,1,2$)
\begin{align}
Q_i&=S_i=K_i, \nn \\
Q_4&=\Big (-\frac{1}{2}+\frac{\sqrt 3}{2}\Big ) K_4, \quad  S_4= \Big (-\frac{1}{2}-\frac{\sqrt 3}{2}\Big ) K_4, \label{reduction1}\\
K_3&=-\dfrac{\sqrt 3}{2} K_4, \quad K_3=K_8, \nn \label{reduction1}
 \end{align}
  which extract three non-trivial one-dimensional irreducible representations $\{\Gamma_1\}_i$ of $S_3$
 acting on field components $(K_i,Q_i,S_i)$ in color space spanned by $\{T^{2,5,7}\}$  and one two-dimensional 
 standard irreducible representation
 $\Gamma_2$  acting  on fields $\{K_4,Q_4,S_4\}$ in the space formed by $\{T^{1,4,6}\}$, the fields 
 $\{K_4,Q_4,S_4\}$ satisfy an equation
 \bea
 K_4+Q_4+S_4=0,  \label{plane}
 \eea
  which defines a two-dimensional plane and implies a vanishing total color charge
  in a similar way as for $I,U,V$-vectors $\Ab_\varphi^p$ in (\ref{SU3DHN}).
 The last two constraints in (\ref{reduction1}) provide consistency with equations of motion.
 With this, the original $SU(3)$ Yang-Mills Lagrangian with gauge fixing terms can be written in an explicit Weyl symmetric form
\begin{eqnarray}
{\cal L}^{\rm Weyl}
\!\!&=&\!\!{\cal L}_0+\sum_{\text{\it {I,U,V}}}{\cal L}_{\rm gf}^{I,U,V}\nn \\
\!\!&=&\!\!\sum_{p}\Big \{  -\dfrac{1}{3} (\pro_\mu \Ab_\nu^p)^2-|D_\mu^p W_\nu^p|^2\nn \\ 
&&\hspace*{1mm}- \dfrac{9}{4} \Big (
 (W^{*p\mu} W^p_\mu)^2-(W^{*p\mu} W^{*p}_\mu) (W^{p\nu} W^p_\nu) \Big )\Big \},  \nn \\
&& \label{Lweyl}
\end{eqnarray}
with 
\begin{align*}
W^{I}_{\mu}&= \dfrac{1}{\sqrt 2} (A_\mu^1 + i A_\mu^2),  & 
W^{U}_{\mu}&=\dfrac{1}{\sqrt 2} (A_\mu^6+ i A_\mu^7), \\
W^{V}_{\mu}&= \dfrac{1}{\sqrt 2} (A_\mu^4-i A_\mu^5), & 
D_\mu^p&=\pro_\mu+i A_\varphi^\alpha r_\alpha^{\,p},
\end{align*}
which coincides with  $SU(2)$ reduced Lagrangian
${\cal L}_{\rm tot}$, (\ref{Lred}), after rescaling  $K_{\hat i} \rightarrow 1/\sqrt 3 K_{\hat i}$.
This implies that Weyl symmetric $SU(3)$ Lagrangian produces the same Euler equations
and solutions as in the case of $SU(2)$ Yang-Mills theory. However,
there is a principal difference: $SU(2)$ solutions are 
degenerated due to the presence of the global color symmetry $SO(2)$ in (\ref{residual}), which causes 
spontaneous color symmetry breaking.
Contrary to this, solutions defined by the ansatz (\ref{SU3DHN},\ref{reduction1}) are non-degenerate 
due to the rigid relationship for the fields $K_{3,4}$, (\ref{reduction1}), which prevents appearance of a global symmetry.

Let us consider eigenvalues of a Lie algebra valued Abelian vector field,
 $\bfA_\mu = (A_\mu^3 T^3,~A_\mu^8 T^8)$, acting in adjoint representation 
 in the Cartan basis. One can find
 \bea 
&&[\bfA_\varphi^p, T_+^p]=K_3 \bfr^p T_+^p,
\eea
where eigenvalues $K_3 \bfr^p$ define color charges of irreducible representations $\{\Gamma_1\}_i$.
The eigenvalues $K_3 \bfr^p$ match the root system with a common field factor $K_3$ and imply 
 zero color charges of representations $\{\Gamma_1\}_i$. With this
a Weyl invariant total 
color charge of the solution $A_\mu^a$ vanishes, as a consequence, all cubic interaction terms are mutually canceled
in the Lagrangian ${\cal L}^{Weyl}$.
All field functions $K_i$ represent fixed points 
under the Weyl transformations, so every Weyl symmetric solution $A_\mu^a$ defined by the 
ansatz (\ref{SU3DHN},\ref{reduction1}) 
represents a fixed point in the configuration space of fields.
Note that, despite on a fact that 
irreducible representation $\Gamma_2$ contains one independent field $K_4$ which is a fixed point under Weyl transformation,
we do not name it as a singlet keeping an accepted terminology in the group theory for  $\Gamma_2$  defined 
as a two-dimensional  representation since it is defined on a two-dimensional plane (\ref{plane}).
In general, a Weyl symmetric Lagrangian admits solutions which are not Weyl symmetric.
Such a case is realized, for example, in one-loop effective Lagrangian with a constant magnetic field background
 \cite{flyvb} where
solutions form a Weyl sextet. In our case each solution transforms into itself in a non-trivial way while $I,U,V$-components of the solution 
permute with each other.
 We constrain our consideration mostly with magnetic type solutions. Similar results 
are valid for electric solutions defined by a dual ansatz in subsection 5.
\subsection{Abelian solutions and Abelian dominance}
A Hilbert space of Abelian Weyl symmetric solutions 
is defined by a complete basis of transverse vector spherical harmonics which represent eigenfunctions
 of a total angular momentum operator with quantum numbers $J=l$, $J_z=m$ \cite{jackson}, 
\be
\begin{array}{rcl}
\vec A_{lm}^{\mathfrak{m}}\!\!\!&=&\!\!\!\dfrac{1}{\sqrt{l(l+1)} }\vec L j_l (k r) Y_{lm} (\theta,\varphi ) 
\, {\rm e}^{i \omega t},\\
\vec A_{lm}^{\mathfrak{e}}\!\!\!&=&\!\!\!\dfrac{-i}{\sqrt{l(l+1)} }\vec \nabla \times (\vec L j_l(k r) Y_{lm} 
(\theta,\varphi))\, {\rm e}^{i \omega t},
\end{array}  
\label{vecharmonics}
\ee
where $j_l(r)$ is a spherical Bessel function, $Y_{lm}(\theta,\varphi)$ is a spherical harmonic, 
$\omega=k\equiv M$ (in units $c=1$) due to conformal invariance, $M$  is a conformal mass scale parameter
 and superscripts $\mathfrak{m,e}$ denote magnetic and electric type, respectively. 
Non-Abelian solutions can be obtained only numerically by solving four partial differential equations and one constraint,
(\ref{eq1})-(\ref{constr1}).
To solve these equations we apply a method which transforms hyperbolic equations (\ref{eq1})-(\ref{constr1}) defined 
on a three-dimensional space-time to a system of $4\times N$ elliptic equations in the two-dimensional space.
First, we decompose fields  $K_{\hat i}(r,\theta,t)$ in a Fourier series
 \be
\begin{array}{rcl}
K_{1,2,4}(r,\theta,t)\!\!&=&\!\!\displaystyle\sum_{n=1}^N \tilde K_{1,2,4}^{(n)}(Mr,\theta)  \cos (n M t), \\
K_0(r,\theta,t)\!\!&=&\!\!\displaystyle\sum_{n=1}^N \tilde K_{0}^{(n)}(Mr,\theta)  \sin(nMt).  
\end{array}
\label{seriesdec}
\ee
After averaging the Lagrangian ${\cal L}_{\rm tot}$ in (\ref{Lred}) over the time period $T=2 \pi/M$, one obtains a 
system of $4\times N$ two-dimensional
equations for Fourier modes $\tK_{\hat i}^{(n)}(r,\theta)$. 
A further simplification
 is achieved by setting all even Fourier modes to zero. This resolves 
the quadratic constraint (\ref{constr1}) and selects a subclass of solutions with a definite parity.

We solve equations in a spherical space domain $\{0\leq r \leq L, 0\leq \theta \leq \pi \}$, 
constrained by radius values $L=\{\mu_{nl},\nu_{nl}\}$  where $\mu_{nl}$,$\nu_{nl}$ are nodes and antinodes
of the spherical Bessel function $rj_l(r)$. We impose the following boundary conditions
\be
\begin{array}{rcl}
\tK_i(r, \theta)^{(n)}\big|_{r=0}\!\!&=&\!\!0, \\ [2\jot]
\tK_i(r, \theta)^{(n)}\big|_{\theta=0,\pi}\!\!&=&\!\!0,
\end{array} 
\ee
and on a spherical boundary one has
\bea
&\begin{cases} \tK_{2,4} (r,\theta)^{(n)}\big|_{r=L}=0,& \mbox{if } L=\mu_{nl} ,
\\   \partial_r \tK_{2,4}(r, \theta)^{(n)}\big|_{r=L}=0,  & \mbox{if }L=\nu_{nl}, \end{cases}\nn \\
&\tK_{0,1}(r,\theta)^{(n)}\big|_{r=L}\approx 0.  \label{conds2}
\eea
In addition, periodic and antiperiodic boundary conditions are used 
for even and odd polar angle modes respectively.
One has to solve a non-linear stationary boundary value problem (BVP). As it is known,
a regular solution to non-linear BVP exists not for arbitrary boundary conditions, and not for arbitrary size of the space domain.
To solve the non-linear BVP we apply iterative numerical methods which generate a convergent solution starting from
approximate  initial profile functions for fields $\tK_i$.Initial profile functions can be found in analytic form in term of series expansion
in local vicinity near $r=0$ and in the asymptotic region $r\rightarrow \infty$.
An advantage of iterative methods is that a final convergent solution 
is not much sensitive to a choice of initial profile functions and values
of integration constants in  boundary conditions.
The initial profile functions for the Abelian field $K_4$ is provided by vector spherical harmonics
with quantum numbers $(l,m)$. Polar angle modes of the 
propagating off-diagonal field $\tK_2(r,\theta)$  are characterized by number $k=0,1,2,\cdots$ of zeros  in the interval $(0\leq \theta \leq\pi)$.
 So one can construct an initial profile for these modes by modifying the Legendre polynomial $P_k(\cos \theta)$.
A numeric solution of magnetic type with the lowest non-trivial polar angle modes is presented in FIG. 1
 in the leading order of Fourier series decomposition which provides sufficiently high accuracy 
 due to structure of the equations defining generalized Jacobi type functions and additional reflection symmetry 
 of solutions at the origin $r=0$ \cite{plb2018}
\be
\begin{array}{rcl}
K_{1,2,4}(r,\theta,t)\!\!\!&=&\!\!\!\tilde K_{1,2,4}(M r,\theta)  \cos (M t),  \\ [2\jot]
K_0(r,\theta,t)\!\!\!&=&\!\!\!\tilde K_0(M r,\theta)  \sin(M t). 
\end{array}
 \label{seriesdec}
\ee
\begin{figure}[h!]
\centering
\subfigure[~]{\includegraphics[width=42mm,height=34mm]{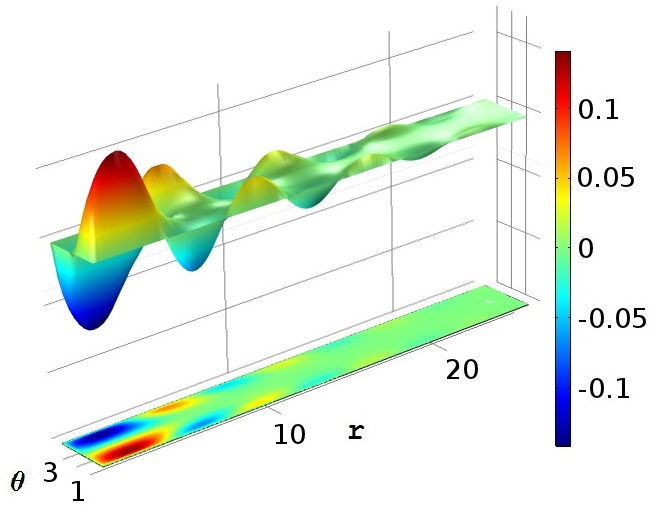}}
\hfill
\subfigure[~]{\includegraphics[width=42mm,height=34mm]{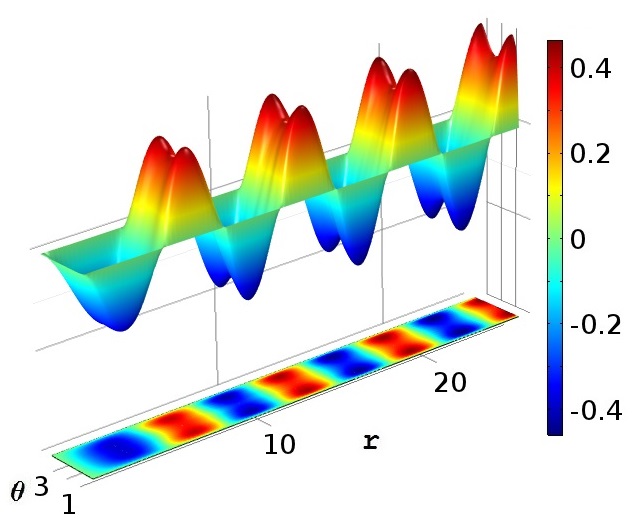}}
\hfill
\subfigure[~]{\includegraphics[width=42mm,height=34mm]{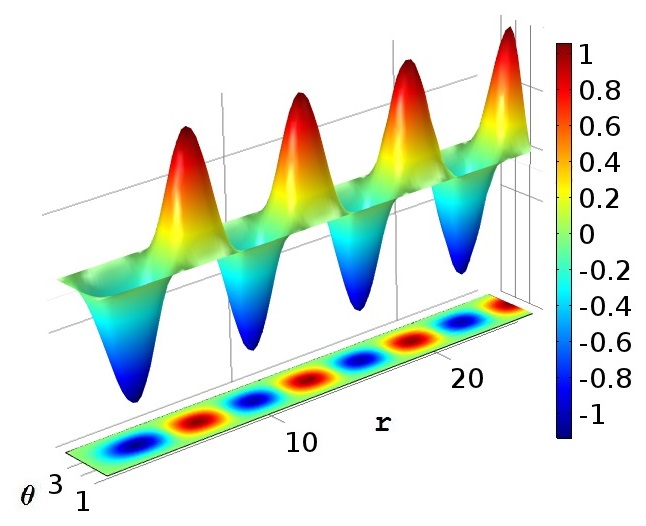}}
\hfill
\subfigure[~]{\includegraphics[width=42mm,height=34mm]{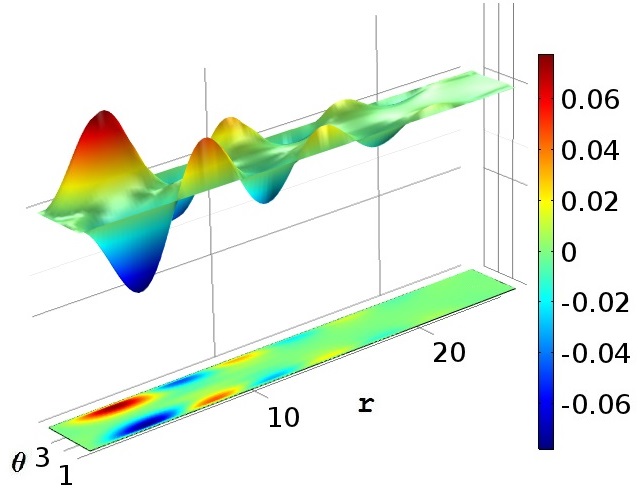}}
\hfill
\subfigure[~]{\includegraphics[width=42mm,height=28mm]{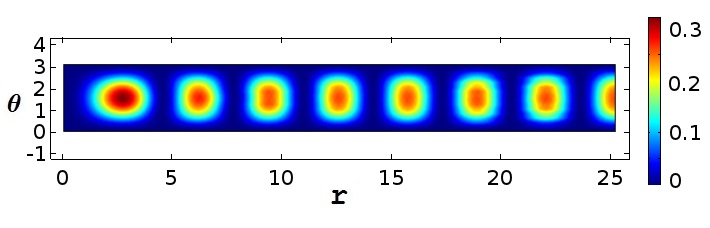}}
\hfill
\subfigure[~]{\includegraphics[width=42mm,height=30mm]{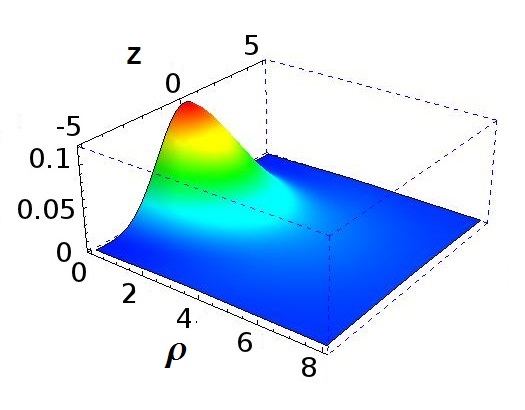}}
\caption[fig1]{Numeric solution in the leading order: (a) $\tK_1$;
(b) $\tK_2$; (c) $\tK_4$; (d) $\tK_0$; (e) time averaged radial magnetic field
$\langle F_{\theta \varphi}^3\rangle_t =-\langle F_{\theta \varphi}^8\rangle_t=-\dfrac{3}{4} \tK_2 \tK_4$;
(f) time averaged energy density $\overline{\!\cal E}(\rho, z)$ 
in cylindrical coordinates 
($g=1,M=1$).}\label{Fig1}
\end{figure}
The Lagrangian ${\cal L}_{\rm tot}$, (\ref{Lred}), does not contain interaction
terms composed from only off-diagonal fields $K_i$. Due to this, non-Abelian solutions exist 
only in the presence of Abelian field $K_3$. This implies an Abelian dominance effect
for low energy solutions.
Indeed, the Abelian numeric profile function $\tK_4(r,\theta)$, FIG. 1(c), coincides with the lowest vector harmonic 
$A_{10}^{\mathfrak{m}}$ with a high accuracy, FIG. 2(a) and Table I.
Moreover, the contribution of the Abelian field
to the total energy in a finite space domain is near $95\% \pm 1.5 \%$, FIG. 2(b),
which is very close to a known estimate of Abelian dominance 
established in the Wilson loop functional \cite{abeldom1,abeldom3}.
 \begin{figure}[h!]
\subfigure[~] {\includegraphics[width=42mm,height=20mm]{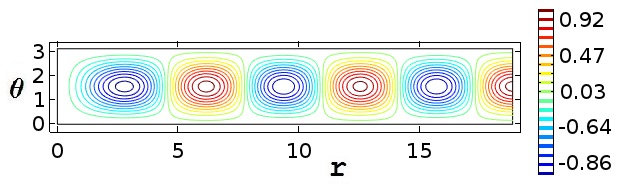}}
\subfigure[~]{\includegraphics[width=42mm,height=22mm]{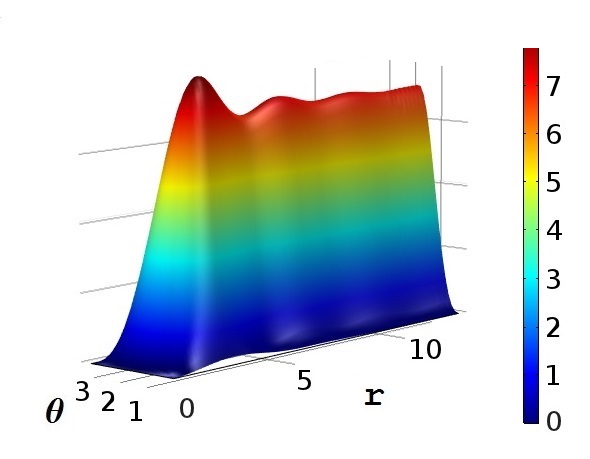}}
\caption[fig2]{(a) A contour plot of the Abelian field $\tK_4$ of the non-Abelian solution, FIG. 1(c);
(b) the time averaged integral energy density $r^2\sin\theta \,{\cal E}$.}\label{Fig2}
\end{figure}
\begin{table}[h]
\begin{center}
\begin{tabular}{|c|c|c||c|c|c|}
\hline
                &           &       &    &    &                    \\
~$\nu_{n1}$  & ~ $\nu_{n1}^{\rm num}$   &  ~$\nu_{n1}^{\rm exact}$ &~ $\mu_{n1} $ & ~
  $\mu_{n1}^{\rm num}$   &~  $\mu_{n1}^{\rm exact}$ \\
\hline
\hline
~$\nu_{11}$ &   2.79  &   2.74 &   $\mu_{11}$&  4.52   &  4.49  \\
\hline
~$\nu_{21}$ &   6.18  &   6.12 &   $\mu_{21}$&  7.81   & 7.73  \\
\hline
~$\nu_{31}$ &   9.37  &  9.32 &   $\mu_{31}$&  10.97   &  10.90  \\
\hline
~$\nu_{41}$ &   12.56  &   12.49 &   $\mu_{41}$&  14.14   & 14.07  \\
\hline
~$\nu_{51}$ &   15.74  &   15.64 &   $\mu_{51}$&  17.29   &  17.22  \\
\hline
\end{tabular} 
\caption{\label{tab:tc} Values of zeros and extremums of the numeric solution $\tK_4$, 
and exact values of nodes and antinodes of the radial part $r j_1(r)$ of the vector harmonic
$\vec A_{10}^{\mathfrak{m}}$.}
\end{center}
\end{table}

\subsection{Structure of the space of non-Abelian magnetic solutions}

 The presence of the Abelian field component in the non-Abelian solution is important since it allows to classify 
 all regular finite energy non-Abelian Weyl symmetric solutions, at least in principle, what is usually not possible in
  non-linear non-integrable theories.
Each Abelian solution defined by a spherical harmonic with quantum numbers  $M$, $J=l$ and $J_z=0$ determines an 
infinite countable set of non-Abelian solutions numerated by number ``$k$'' of zeros of the polar angle mode of the field $\tK_2$.
Setting approximate initial profile functions for $K_i$  and appropriate boundary conditions providing finite energy density 
a numeric solution is defined uniquely and can be obtained by using iterative numerical methods.
Solution with the lowest non-trivial polar angle mode of the field $\tK_2$ is shown in FIG. 1. 
The next solution with $l=1, k=1$ is depicted in FIG. 3. The solution has  the same energy density function
and an opposite parity
of the off-diagonal field $\tK_2$ compare to the solution with quantum numbers $l=1,k=0$.
\begin{figure}[h!]
\centering
\subfigure[~]{\includegraphics[width=42mm,height=30mm]{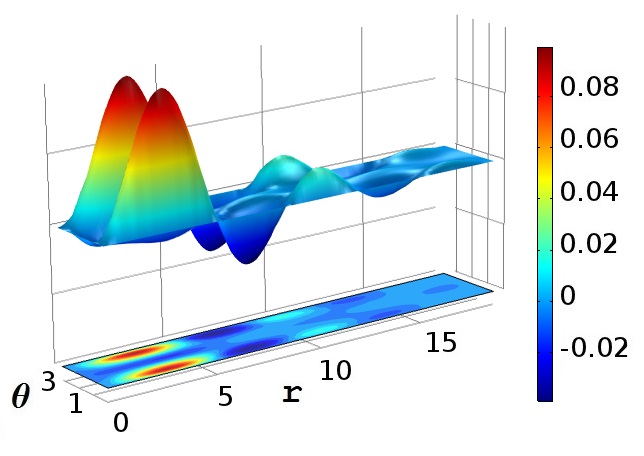}}
\hfill
\subfigure[~]{\includegraphics[width=42mm,height=30mm]{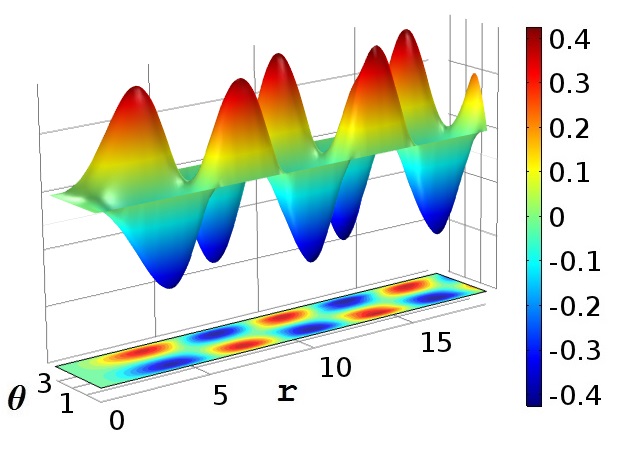}}
\hfill
\subfigure[~]{\includegraphics[width=42mm,height=30mm]{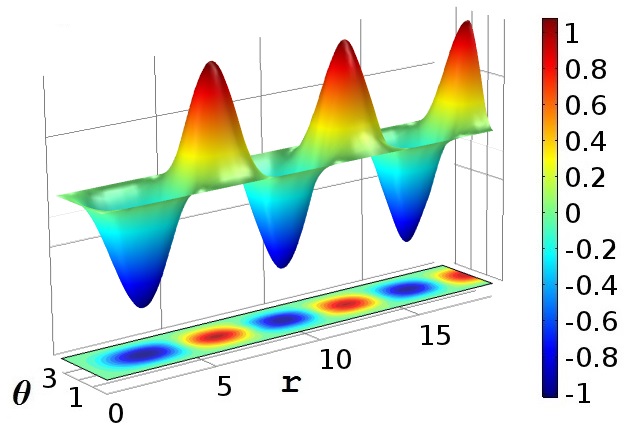}}
\hfill
\subfigure[~]{\includegraphics[width=42mm,height=30mm]{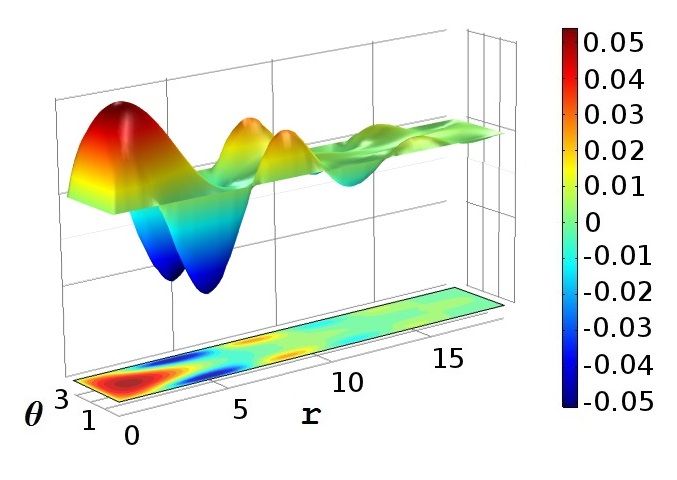}}
\subfigure[~]{\includegraphics[width=50mm,height=38mm]{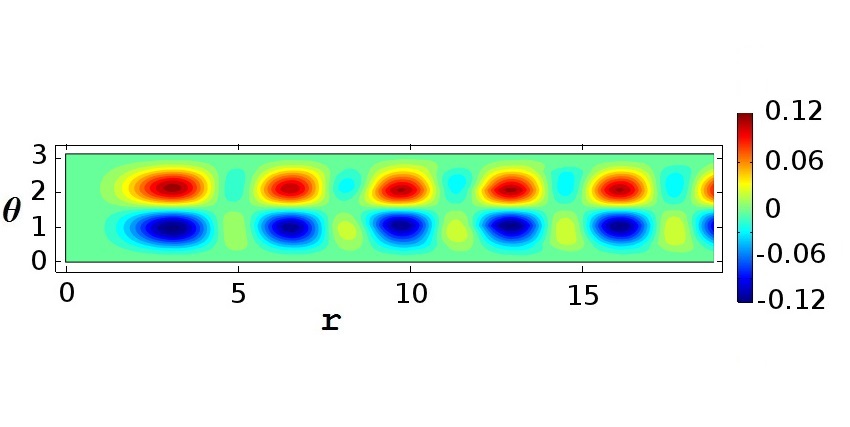}}
\hfill
\caption[fig3]{Solution profile functions  with quantum numbers $l=1,k=1$ in spherical coordinates: (a) $\tK_1$;
(b)  $\tK_2$; (c)  $\tK_4$; (d)   $\tK_0$;
(e) the time averaged radial color magnetic field $\overline { B_r^3}=-3/4 \tK_2 \tK_4$; 
($g=1,M=1$).}\label{Fig3}
\end{figure}
To show clearly the location of zeros of the polar mode of $\tK_2$ we present density plots for higher 
mode solutions with $l=1, k=2,3$ in FIGs. 4,5.
 \begin{figure}[h!]
\centering
\subfigure[~]{\includegraphics[width=42mm,height=30mm]{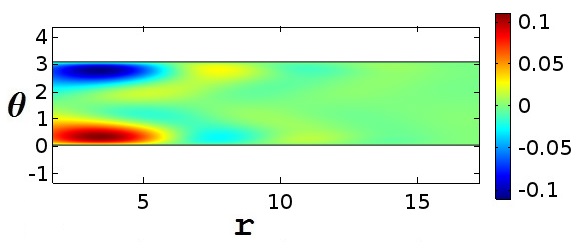}}
\hfill
\subfigure[~]{\includegraphics[width=42mm,height=30mm]{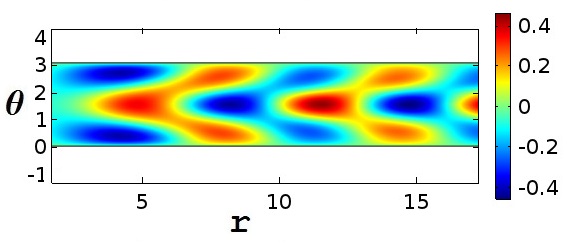}}
\hfill
\subfigure[~]{\includegraphics[width=42mm,height=30mm]{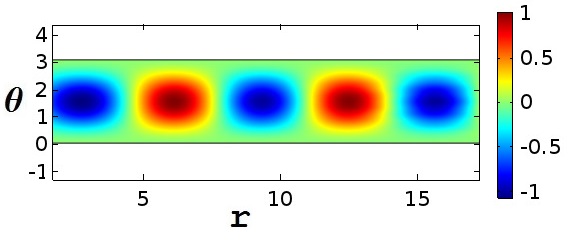}}
\hfill
\subfigure[~]{\includegraphics[width=42mm,height=30mm]{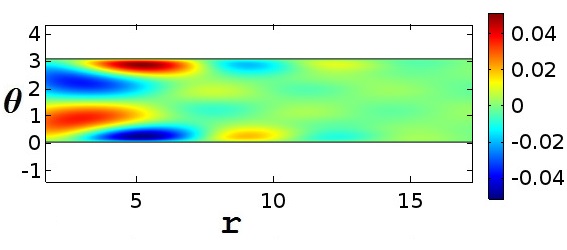}}
\hfill
\caption[fig4]{Solution with quantum numbers $l=1,k=2$ in spherical coordinates: (a) $\tK_1$;
(b)  $\tK_2$; (c)  $\tK_4$; (d)   $\tK_0$ ($g=1,M=1$).}\label{Fig4}
\end{figure}
 \begin{figure}[h!]
\centering
\subfigure[~]{\includegraphics[width=42mm,height=30mm]{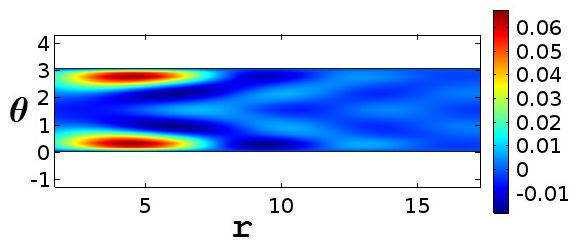}}
\hfill
\subfigure[~]{\includegraphics[width=42mm,height=30mm]{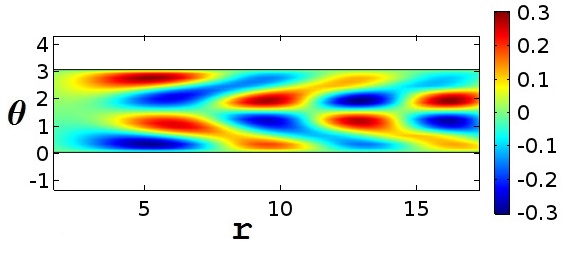}}
\hfill
\subfigure[~]{\includegraphics[width=42mm,height=30mm]{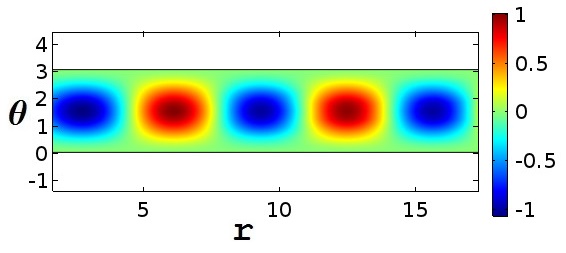}}
\hfill
\subfigure[~]{\includegraphics[width=42mm,height=30mm]{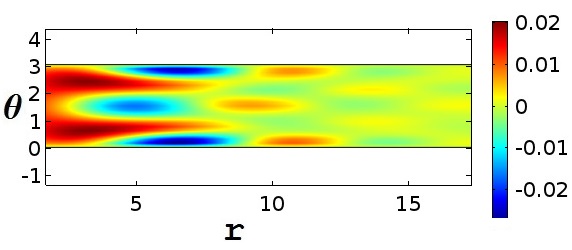}}
\hfill
\caption[fig5]{Solution  with quantum numbers C in spherical coordinates: (a) $\tK_1$;
(b)  $\tK_2$; (c)  $\tK_4$; (d)   $\tK_0$ ($g=1,M=1$).}\label{Fig5}
\end{figure}
One can observe, that lowest mode solution in FIG. 1 contains a non-vanishing radial component 
of the magnetic field $\langle F_{\theta \varphi}^3\rangle_t =-\dfrac{3}{4} \tK_2 \tK_4$ of Coulomb type, FIG. 1(e),
which can be treated as a magnetic field of the magnetic stationary monopole.
The solution with quantum numbers $(l=1,k=1)$
has non-vanishing radial magnetic field of higher mode, FIG. 3(e), which can be treated 
as a field of a monopole-antimonopole pair.
 We have solved equations in space domain with radial  size in the interval $\nu_{11}\leq L\leq 200$, 
 all solutions show fast convergence with number of iterations less than $10$.

The most intriguing question is: ``what is the dimension of the solution space defined by a given set of
quantum numbers $(M,l,m,k)$?'' Due to the presence of local $U(1)$ gauge symmetry (\ref{residual}),
one would expect that there are two dynamic degrees of freedom corresponding to two long-distance
propagating modes $K_2, K_4$. On the other hand, due to the Abelian dominance effect it is clear that off-diagonal 
components can not represent independent dynamic degrees of freedom due to non-linear interaction between 
fields $K_{\hat i}$. Moreover, in the limit $K_4\rightarrow 0$ the equations for off-diagonal fields $K_i$ 
turn precisely into Maxwell  equations for electric type vector harmonics. However, the 
magnetic ansatz does not describe non-Abelian electric dual solutions, 
since the radial non-Abelian electric field strength vanishes identically within the magnetic ansatz.
A dual electric ansatz is presented in the next subsection. Therefore, off-diagonal field $\tK_2$ must be correlated 
with the Abelian field $\tK_4$.
Indeed, it is surprising, a careful numeric analysis of solutions with quantum numbers $l=1, k=0,1,2,3$ 
shows that amplitudes of the fields $K_i$ are correlated with the amplitude of the Abelian field: once fixed the 
value of the asymptotic amplitude of the Abelian field, the amplitudes of other 
modes $K_{0,1,2}$ are uniquely determined by the numeric procedure irrespective of initial amplitude values of 
$K_{0,1,2}$. A source of such a non-trivial feature is related to the fact  that 
spherical harmonics can describe localized states in specific space domains 
constrained by a sphere with selected radius values, $R=\{\nu_{nl},\mu_{nl}\}$, since only in these cases
 the total energy inside space domain is conserved and classically stable stationary solutions exist. 
 In addition, for small amplitudes of the Abelian field $K_3$ the field profiles  $K_{1,2}$ are given approximately by
the electric vector harmonic with the same spherical Bessel function $j_l(r)$.
So that the radial parts of modes $K_{2,4}$ in the non-Abelian solution must have the same,
at least one node or antinode to provide energy conservation inside finite space domain. This causes correlation 
between amplitudes of fields $K_{2,4}$ since for  non-Abelian solutions the location of nodes/antinodes depend on field
amplitudes.This is demonstrated in a simplified  $1+1$ dimensional model obtained by averaging equations for $K_{2,4}$ 
over time period and polar angle in Appendix FIG. 10(a),(b). 

We conclude {\it  Weyl symmetric solutions defined by ansatz (\ref{SU3DHN},\ref{reduction1})
have a total color charge zero, and a space of solutions 
for a given set of quantum numbers $(M,l,m,k)$ is one-dimensional and defined by one normalization 
constant, an asymptotic amplitude of the Abelian field. This lead to color singlet one-particle quantum states after quantization.}

\subsection{Duality, Weyl symmetric ansatz \\for electric solutions}

Non-Abelian field strength components $F_{\mu\nu}^a$
are not gauge invariants like the magnetic and electric fields in the Maxwell theory. 
So that, the duality symmetry between non-Abelian magnetic and electric
solutions can not be realized as a symmetry with respect to mutual exchange of color electric and magnetic 
fields $\vec B^a \rightarrow \pm \vec E^a$. We define dual non-Abelian electric and magnetic fields by imposing
minimal requirements that: (i) dual fields have the same energy density
function;(ii) the gauge invariants $(F_{\mu\nu}^a)^2$ corresponding to vacuum gluon condensates must be equal by module 
and opposite by sign; (iii) magnetic non-Abelian solution contains a non-vanishing time averaged radial magnetic field 
$\langle F_{\theta \varphi}^3\rangle_t$ of Coulomb type, and a dual electric field contains time averaged non-vanishing 
radial electric field $\langle F_{t r}^3\rangle_t$ of Coulomb type.

With this one can construct $SU(3)$ Weyl symmetric ansatz for dual electric fields.
A minimal axially symmetric ansatz contains the following non-vanishing 
components of the gauge potential 
\bea
A_{r,\theta,\varphi}^{3,8}\!\!&=&\!\!-\dfrac{\sqrt 3}{2} K_{1,2,3},   \nn \\
A_{r,\theta,\varphi}^1\!\!&=&\!\!K_{1,2,3},      \nn \\
A_{r,\theta,\varphi}^4\!\!&=&\!\!\big (-\dfrac{1}{2}+\dfrac {\sqrt 3}{2}\big)K_{1,2,3},  \label{ElAns}\\
A_{r,\theta,\varphi}^6\!\!&=&\!\!\big (-\dfrac{1}{2}-\dfrac {\sqrt 3}{2}\big)K_{1,2,3},  \nn \\
A_t^2&=&K_0, \quad A_t^5=-K_0, \quad A_t^7=K_0.   \nn
\eea
To solve equations of motion it is suitable to remove redundant gauge symmetry and pure gauge
degrees of freedom. So one should impose gauge conditions in such a way that the Weyl symmetry
survives and equations for independent fields $K_i$ keep consistence with equations for the full
gauge potential and with Maxwell equations in the Abelian limit. In a case of the electric ansatz 
(\ref{ElAns}) all requirements are satisfied if we introduce gauge fixing terms corresponding to
the covariant Lorenz gauge
\be
{\cal L}_{\rm gf}=-\dfrac{1}{2}\sum_{a} \big (-\pro_t A_t^a+({\rm div} \vec A)^a\big)^2,
\ee
which differs from the gauge fixing term for magnetic solutions (\ref{gfterm}).
A Weyl symmetric electric type solution with lowest non-trivial polar modes is presented in FIG. 6
 in the leading order of Fourier series decomposition.
 \begin{figure}[h!]
\centering
\subfigure[~]{\includegraphics[width=42mm,height=30mm]{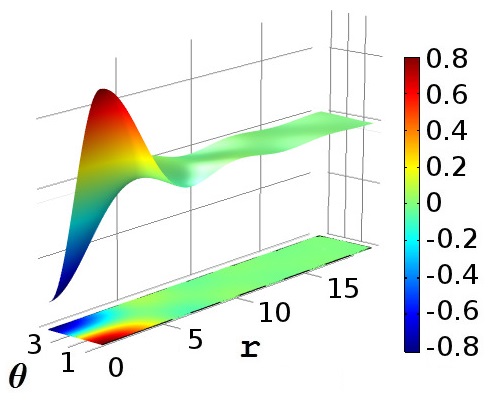}}
\hfill
\subfigure[~]{\includegraphics[width=42mm,height=30mm]{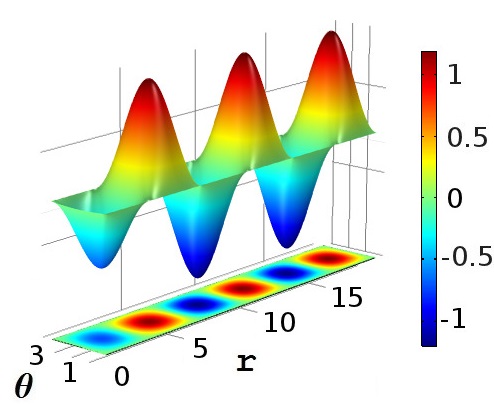}}
\hfill
\subfigure[~]{\includegraphics[width=42mm,height=30mm]{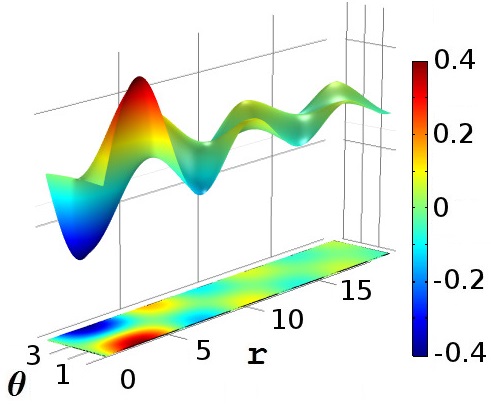}}
\hfill
\subfigure[~]{\includegraphics[width=42mm,height=30mm]{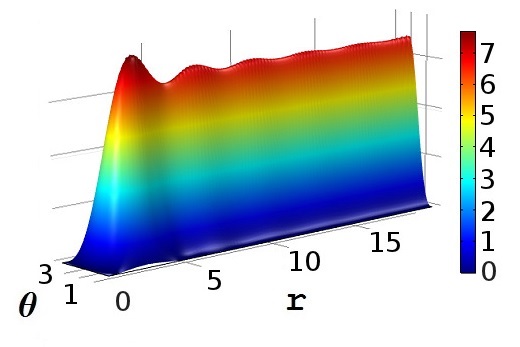}}
\caption[fig6]{Solution profile functions of electric solution: (a) $\tK_1$;
(b)  $\tK_2$; (c)  $\tK_0$; (d)  The time averaged energy density
($g=1,M=1$).}\label{Fig6}
\end{figure}
The solution has the same energy density profile function as a magnetic solution,
FIG. 2(b), and the vacuum condensate function $(F_{\mu\nu}^a)^2$
has an opposite sign to compare with the magnetic solution, as it should be for a
dual electric solution. 
Note that, even though gauge field components $K_{1,0}$ take non-zero values along the $Z$-axis,
and $K_1$ is a non-vanishing function at the origin $r=0$, all field strengths and energy density
are regular single valued functions everywhere. The magnetic potential $K_3$ vanishes identically, 
this indicates that space of electric solutions defined by the ansatz (\ref{ElAns}) 
is one dimensional and leads to color-singlet one-particle quantum states after quantization
the same way as for magnetic solutions.\\

\section{Weyl symmetric Dirac fermions}

 Now we consider an effect of the Weyl symmetry on matter fields described by Lagrangian ${\cal L}_{\rm f}$
for fermions
\bea 
{\cal L}_{\rm f}= \bar \Psi \Big[ i \gamma^\mu (\pro_\mu -\dfrac{i g}{2} A_\mu^a \lambda^a)-m\Big]\Psi.
\eea
The Euler equations to the total Yang-Mills Lagrangian ${\cal L}_{\rm tot}={\cal L}_0+{\cal L}_{\rm f}$ read
\bea
&&(D^\mu \vec F_{\mu\nu})^a= -\dfrac{g}{2} j^a \equiv -\dfrac{g}{2}  \bar \Psi \gamma_\nu \lambda^a\Psi, \label{eqA}\\
&& \Big[ i \gamma^\mu (\pro_\mu -\dfrac{i g}{2} A_\mu^a \lambda^a)-m\Big]\Psi=0. \label{eqPsi}
\eea
A usual simple Abelian projection with  two independent Abelian fields $A_\mu^{3,8}$ corresponding to 
the Cartan generators leads to Dirac equations for three independent color quarks 
\bea
   \Big[ i \gamma^\mu \pro_\mu -\dfrac{i g}{2}  \sum_p A_\mu^\alpha w_\alpha^p-m\Big]\Psi_p=0,
\eea
where $w_\alpha^p$ are the weight vectors $\bfw^p=\{(1, 1/\sqrt 3), (-1,1/\sqrt 3), (0,-2/\sqrt 3)\}$.
The equations have three independent solutions for quarks forming a color triplet.
Note that the simple Abelian projection is not consistent with Weyl symmetric structure of the full Lagrangian
including off-diagonal gluon fields. 
An important feature of the Weyl symmetric ansatz (\ref{SU3DHN},\ref{reduction1})
 is that it implies a non-trivial Abelian projection
 with one type Abelian field $K_3$ located in extended  color subspace spanned by
 generators $(T^{3,8,1,4,6})$. Applying the ansatz to (\ref{eqPsi}) results in 
different equations for quarks
\be
 \Big[ i \gamma^\mu \pro_\mu-m+\dfrac{g}{2} \gamma^\mu  A_\mu G 
                +\dfrac{g}{2}\gamma^\mu K_\mu Q\Big]\Psi=0, \label{GQeqns} 
\ee
with
\be
  G=\begin{pmatrix}
   \tw^1&\tw^3&\tw^2\\
  \tw^3&\tw^2&\tw^1\\
   \tw^2&\tw^1&\tw^3\\
 \end{pmatrix},
\qquad
 Q=\begin{pmatrix}
   0&-i & i\\
  i &0&-i\\
   -i&i&0\\
 \end{pmatrix}, 
\label{Gmatrix}
 \ee
where $A_\mu=\delta_{\mu \varphi}K_3$,  $K_\mu=\delta_{\mu \hat n} K_{\hat n}$($\hat n=0,1,2$) and  
$G$ is a color charge matrix  composed from weights, $\tw^p=w^p_3+w^p_8$.  
The equations  (\ref{eqA}) reduce to four independent equations for the fields $K_\mu$. 
 The charge matrix $G$ has three eigenvectors $u^{0,\pm}$ corresponding to eigenvalues
$\lambda^{0,\pm}$
\begin{align}
\lambda^0&=0,~~\lambda^{\pm}=\pm \sqrt{\tg},\nn \\
u^0&= 
\begin{pmatrix}
   1\\
 1\\
  1\\
 \end{pmatrix},~u^\pm=\begin{pmatrix}
   \tw^1\tw^3+\tw^2(\pm\tg-\tw^2)\\
 \tw^2\tw^3+\tw^1(\pm\tg-\tw^1))\\
   \tw^1\tw^2+\tw^3(\pm\tg-\tw^3)+\tg^2\\
 \end{pmatrix},\label{tg} \\
\tg^2&= (\tw^1)^2+(\tw^2)^2+(\tw^3)^2-\tw^1\tw^2-\tw^2\tw^3-\tw^3 \tw^1,\nn
\end{align}
where $\tg$ is a Weyl invariant color charge.
 It is surprising, the system of non-linear coupled equations  (\ref{eqA}),(\ref{eqPsi}) 
admits a free color singlet solution $\Psi^0=\psi^0(x)u^0$ which belongs to one-dimensional irreducible representation 
$\Gamma_1$ of the Weyl group and satisfies a free Dirac equation 
 \bea
&&   \big ( i \gamma^\mu \pro_\mu-m \big) \Psi^0=0.
\eea
The solution implies a vanished color current $j^a$ and full decoupling of equations for gluons and quarks.
So that a complete basis of solutions describing the free gluon and quark solutions is given 
by vector and spinor spherical harmonics with the same quantum numbers $J,m, \mu_{nl}/\nu_{nl}$
which provide the total energy conservation in a finite space domain. The zero mode solutions describe non-interacting 
vacuum gluon and quark condensates and primary gluons and quarks corresponding to vacuum excitations.
Solutions $\Psi^\pm=\psi^\pm(x) u^\pm$  corresponding to eigenvalues $\lambda^\pm$ belong to a standard two-dimensional 
representation $\Gamma_2$  and have a total zero color charge due to equation $\Psi_1+\Psi_2+\Psi_3=0$ defining $\Gamma_2$. 
The coordinate part $\psi^\pm$
satisfies a Dirac equation with a Weyl invariant color charge $\tg$ which represents
an effective coupling constant $\tg=\sqrt 6$,  
\be
\big ( i \gamma^\mu \pro_\mu-m \big)\pm g\tg \gamma^\mu A_\mu^3 \big) \psi^\pm=0,
 \ee
  Solutions $\Psi^{0\pm}$ 
  form three one-dimensional invariant Weyl symmetric subspaces and represent an 
  orthogonal basis in the vector space of the representation $\Gamma_2\oplus \Gamma_1$.  
   It is clear that after quantization such solutions lead to primary color singlet one-particle quantum states
 for quarks. 
  It is remarkable,{\it the Weyl symmetric solutions describe only color singlet quantum states for gluons and quarks,
 contrary to commonly accepted wisdom on color nature of gluons and quarks. Color gluons and quarks can not be observed
 because their concepts had been defined in past on a basis of the perturbation theory and applied a simple Abelian projection 
 which is inconsistent with the whole non-Abelian Weyl symmetric structure (\ref{SU3DHN},\ref{reduction1}) of the 
 full Yang-Mills Lagrangian}. Certainly, 
 some special solutions, like ordinary Abelian plane wave or 
 non-linear plane waves, exist, however, such solutions are not physical due to another important condition for 
 physical vacuum and states - quantum stability against vacuum fluctuations.

\section{Quantum stable vacuum}

The vacuum stability has been a persistent problem in all vacuum models in QCD since 1977 \cite{savv,N-O}.
We outline in short the main difficulties on the way of constructing a true stable vacuum, and then
we prove explicitly the quantum stability of Abelian vector harmonics by solving
a system of eigenvalue equations for unstable (tachyonic) modes.
The quantum instability of constant color magnetic QCD vacuum was established in \cite{N-O}, 
and later it was found for constant color electric background as well \cite{schan}.
The source of quantum instability is the presence of anomalous magnetic moment interaction
terms in the classical Yang-Mills Lagrangian which implies that one-loop effective action
gains an imaginary part, i.e., vacuum is unstable. Numerous attempts to construct a stable vacuum
from static field configurations fail in the class of regular solutions in Minkowski space-time.
It was noticed that at small space-time scale the elementary field configurations 
should be vibrating due to quantum mechanical principle \cite{niel-oles2,amb-oles2}.
A simple analysis of plane wave solutions shows the presence of quantum instability \cite{prd2017},
since the eigenvalue problem for unstable modes represents a quantum mechanical bound state
problem in one-dimensional periodic potential which admits bound states for any shallow potential well.
It is clear, that stationary solution like a spherical wave in three-dimensional space remove such
obstacle and can provide a stable vacuum configuration. Indeed, a spherically symmetric stationary 
wave solution possesses quantum stability \cite{ijmp2017, ptep2018}.  Unfortunately, the spherical wave 
solution has classical instability and can not describe a physical state. So, an axially-symmetric non-linear 
stationary solution of Yang-Mills theory has been proposed as a stable vacuum field
in \cite{plb2018}. It is very difficult to prove rigorously classical stability of such solution, however,
later it was realized that in the Abelian limit the solutions turn into the known vector spherical harmonics
which represent free photons, which are obviously classically stable since they represent a free gas.
An important role in providing the deepest vacuum is played by Weyl symmetry which selects
Abelian color singlet solution due to the constraint on Abelian fields $A_\mu^3=A_\mu^8$.

Weyl symmetric solutions have a vanished total color charge which implies
mutual cancellation of all cubic interaction terms in the Lagrangian
${\cal L}_{\rm red}$, (\ref{Lred}). 
As it is known, the cubic interaction terms correspond to anomalous magnetic moment interaction which
causes the vacuum instability. The absence of such terms 
plays a principal role in providing quantum stability of Weyl symmetric solutions. 
Quantum stability of the lowest energy non-Abelian Weyl symmetric solution 
has been proved numerically in \cite{plb2018}. 
We demonstrate quantum stability of vacuum Abelian solutions localized in a finite
space region.
We solve a ``Schr\"{o}dinger'' type eigenvalue equation for
quantum gluon fluctuations $\Psi_\mu^a$ 
\begin{align}
\Ka_{\mu\nu}^{ab} \Psi_\nu^b&=\Lambda \Psi_\mu^a,  \label{schr3} \\
\Ka_{\mu\nu}^{ab}&=-
\delta^{ab} \delta_{\mu\nu} \pro^2_t
            -\delta_{\mu\nu}({\Da}_\rho {\Da}^\rho)^{ab} -2 f^{acb}{\mathcal
	    F}_{\mu\nu}^c,  \nn
\end{align}
where the operator $\Ka_{\mu\nu}^{ab} $ corresponds to the gluon contribution 
to one-loop effective action \cite{plb2018}, and $\Da_\mu, \mcF_{\mu\nu}$ 
are defined by means of a background vector harmonic (\ref{vecharmonics}).  
The existence of a negative eigenvalue $\Lambda$
would imply the vacuum instability.
An analysis of vacuum quantum stability is similar to one 
performed in \cite{plb2018}. 
Due to the presence of Weyl symmetry the vacuum stability problem in $SU(3)$
QCD reduces to a corresponding problem in $SU(2)$ theory.
An explicit vector harmonic for axially-symmetric Abelian solution of magnetic type
reads
\begin{align}
B_\mu^a&=\delta_{\mu 3} \delta^{a,3} c_0 r j_l(M r) \sin^2 \theta \cos (M t)\nn \\
   &\equiv \delta_{\mu 3} \delta^{a,3} B_\varphi(r,\theta,t). \label{Abelsol}
\end{align}
Substitution of the solution $B_\mu^a$ into the eigenvalue equations (\ref{schr3}) 
leads to factorization of the initial twelve equations to three systems
of equations for three sets of fluctuating fields: (I) $\{\Psi_1^1,\Psi_2^1,\Psi_3^2, \Psi_0^1\}$,
(II) $\{\Psi_1^2,\Psi_2^2,\Psi_3^1,\Psi_0^2\}$, (III) $\{\Psi_\mu^3\}$. 
The equations from group (III) represent free equations and does not
admit solutions with negative eigenvalues. The second system of equations is equivalent to the first
one after changing variables $\Psi_1^2\rightarrow \Psi_1^1, \Psi_2^2 \rightarrow \Psi_2^1,
\Psi_3^1 \rightarrow \Psi_3^2, \Psi_0^2 \rightarrow \Psi_0^1$ and reflection of the 
background field, $B_\varphi\rightarrow -B_\varphi$.
So one has to solve only one system of four eigenvalue equations
\begin{align}
\Delta \Psi_1^1+\dfrac{1}{r^2} \Big ((2+\csc^2 \theta B_\varphi^2)\Psi_1^1+2(\cot\theta+
 \pro_\theta )\Psi_2^1& \nn\\
-2 \csc \theta (B_\varphi-r\pro_r B_\varphi)\Psi_3^2 &\Big )=\lambda \Psi_1^1, 
\\
\Delta \Psi_2^1 +\dfrac{1}{r^2} \Big (\csc^2\theta(1+B_\varphi^2) \Psi_2^1-
2 \pro_\theta \Psi_1^1&\nn \\
-2\csc\theta(\cot \theta B_\varphi-\pro_\theta B_\varphi) \Psi_3^2&\Big )=\lambda \Psi_2^1,
\\
\Delta \Psi_3^2 +\dfrac{1}{r^2} \Big ((1+\cot^2\theta+\csc^2\theta B_\varphi^2)\Psi_3^2
& \nn \\
-2\csc \theta (B_\varphi-r\pro_r B_\varphi) \Psi_1^1& \nn \\
-2\csc \theta (\cot\theta B_\varphi-\pro_\theta B_\varphi) \Psi_2^1 &\Big )
=
\lambda \Psi_3^2,  
\\
\Delta \Psi_0^1 +\dfrac{1}{r^2 \sin^2 \theta} B_\varphi \Psi_0^1 +
\dfrac{2}{r \sin \theta} \pro_t B_\varphi \Psi_3^2&=\lambda \Psi_0^1, \label{4Psi} 
\end{align}
with
$$
\Delta \equiv-\Big ( \pro^2_t+\pro^2_r +\dfrac{2}{r}\pro_r +\dfrac{1}{r^2} \pro^2_\theta+
\dfrac{\cot \theta}{r^2} \pro_\theta \Big ), 
$$
%
where $\Delta$ is a common part of the vector Laplace operator acting on the gluon fluctuation function.
\begin{figure}[h!]
\centering
\subfigure[~]{\includegraphics[width=42mm,height=32mm]{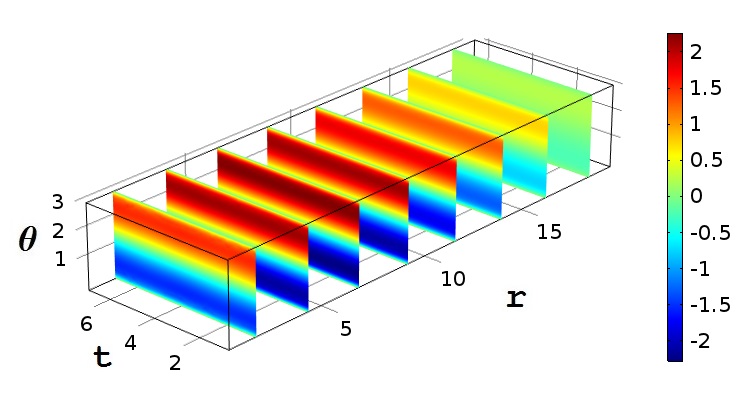}}
\hfill
\subfigure[~]{\includegraphics[width=42mm,height=32mm]{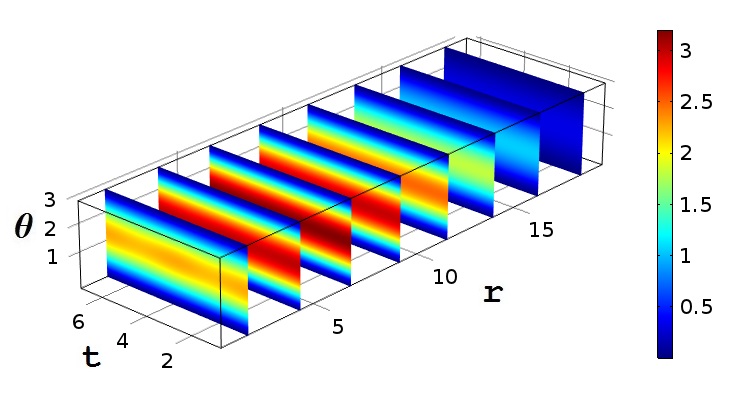}} \\
\subfigure[~]{\includegraphics[width=42mm,height=32mm]{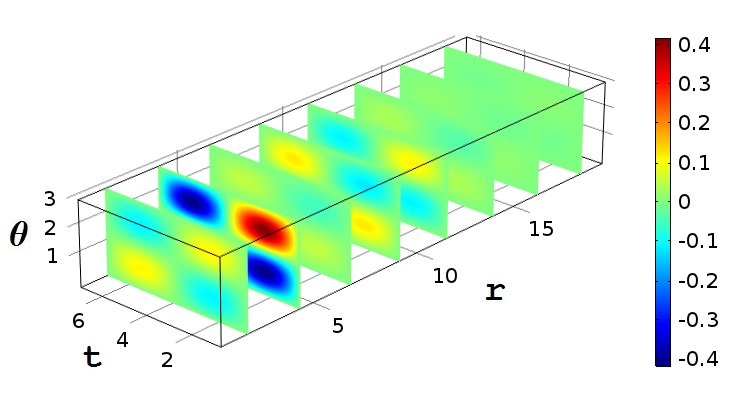}}
\hfill
\subfigure[~]{\includegraphics[width=42mm,height=32mm]{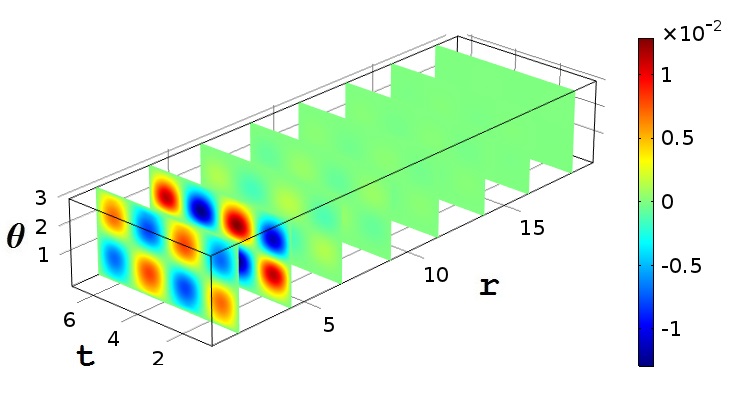}}
\caption[fig7]{Solution for fluctuation modes corresponding to the lowest
eigenvalue $\lambda=0.0292$: (a) $\Psi_1^1$; 
 (b) $\Psi_2^1$; (c) $\Psi_3^2$; (d) $\Psi_0^1$ ($g=1,M=1,c_0=1,l=1$).
}\label{Fig7}
\end{figure}
The obtained numeric solutions with the lowest eigenvalues
confirm the absence of negative modes in the case of parameter values 
$(l \leq 4;\, c_0 \leq 2; \, M\leq 10)$ for size of the space domain $\mu_{11}\leq L\leq 100$, FIG. 7.
Quantum stability of Abelian solutions for finite size space domains constrained by
lowest nodes and antinodes has been demonstrated in (\cite{PRL1}, FIG. 2).

\section{Localization of quantum states and formation of lightest glueballs}

We quantize the Abelian solutions (\ref{vecharmonics}) in a finite space 
region constrained by a sphere of radius $a_0$ corresponding to an effective
glueball size.
It is suitable to introduce dimensionless units $\tilde M=M a_0, x=r/a_0, \tau=t/a_0$.
To find proper boundary conditions we require that the Pointing vector 
$\vec {\mathbf S}=\vec {\mathbf E} \times \vec {\mathbf B}$  
vanishes on the sphere. This implies two possible types of boundary
conditions
\be
\begin{array}{rcrclcrcl}
({\rm I}): &\quad& \vec A_{lm}^{\mathfrak{m,e}} (\tilde M x)\big|_{x=1}\!\!&=&\!\!0, &\quad& \tilde M_{nl}\!\!&=&\!\!\mu_{nl},  \\ [2\jot]
({\rm II}): &\quad& \pro_r(r \vec A_{lm}^{\mathfrak{m,e}} (\tilde M x))\big|_{x=1}\!\!&=&\!\!0, &\quad& \tilde M_{nl}\!\!&=&\!\!\nu_{nl}, 
\end{array}
\label{munu}
\ee
where $\tM_{nl}$ stands for nodes $\mu_{nl}$ and antinodes $\nu_{nl}$ of the Bessel function $ j_l(r)$.
We choose the following normalization condition for the
vector harmonics $\vec A_{lm}^{\mathfrak{m, e}}(\tM x)$ 
\bea
\dfrac{1}{4\pi}\int_0^1\!\!{\rm d}x \!\!\int\!{\rm d}\theta\, {\rm d}\varphi\,  x^2 \sin\theta(\{\vec A_{lm}^{\mathfrak{m, e}}(\tM x))^2
=\dfrac{1}{\tM_{nl}}. \label{normcond}
\eea
The standard canonical quantization results in the following Hamiltonian
expressed in terms of the creation and annihilation operators   $c^{\pm}_{nl}$
\bea
H= \dfrac{1}{2} \sum_{n,l,m} \tilde M_{nl} ( c^+_{nlm} c^-_{nlm}+ c^-_{nlm} c^+_{nlm}). \label{Hamvac}
\eea
One particle states $\{c^+_{nlm} |0\rangle\}$ describe free primary gluons
which are not observable quantities since we have not taken into account 
their interaction to vacuum gluon condensate. 
We follow an idea that vacuum gluon and quark condensates represent inevitable attributes of 
hadrons \cite{brodsky2012}. 

We apply a simple model based on one-loop effective Lagrangian of QCD which
describes appearance of localized solutions corresponding to the lightest glueballs.
Certainly, one loop effective potential is not a much appropriate tool for quantitative  description of glueballs,
nevertheless, it contains a non-perturbative part originated from summation of contributions from infinite number
of one-loop quantum corrections. This provides qualitative description of formation of glueballs as a result of interaction
of primary gluon with corresponding generated vacuum gluon condensates.

We apply one-loop effective Lagrangain for $SU(3)$ pure QCD with Abelian background field
corresponding to Abelian projection defined by the ansatz (\ref{SU3DHN}, \ref{reduction1}).
In approximation of slowly varied fields one can calculate a one-loop effective Lagrangian
\bea
{\cal L}^{\scriptsize\mbox{1-l}} =-\dfrac{1}{4} \tF^2-k_0 g^2 \tF^2 \bigg(\log \Big(\dfrac{g^2 \tF^2}{\Lambda_{\rm QCD}^4}\Big)-c_0 \bigg), 
\label{Leff}
\eea
where the number factor $k_0$ in front of logarithmic term is three times larger to compare with the effective Lagrangian 
for $SU(3)$ Yang-Mills theory with gluons in adjoint representation and quarks in fundamental representation.
The factor three appears due to contribution of Weyl symmetric $I,U,V$ type Abelian gluon fields. Contribution of 
quarks is proportional to sum of squared charges  of three quarks which provides the same factor three due to the value 
of the effective color charge $\tg=\sqrt 6$. So that the number coefficeint in front 
of logarithmic term is still proportional to the standard beta function of $SU(3)$ QCD with the same critical number of flavor quarks.
The field $\tF_{\mu\nu}$  is an external Weyl symmetric Abelian color magnetic field, corresponding to Abelian projection
defined by Weyl symmetric ansatz. 
We will treat the parameters $k_0, c_0$ as free model parameters.
The most important property of the effective Lagrangian is the presence of 
the non-perturbative logarithmic term which generates  a non-trivial minimum
of the effective potential $V^{\scriptsize\mbox{1-l}}=-{\cal L}^{\scriptsize\mbox{1-l}}$ at non-zero value of the vacuum condensate \cite{savv}
\bea
g^2 B_{\mu\nu}^2 =\Lambda_{\rm QCD} \exp \Big(c_0-1-\dfrac{1}{2k_0g^2}\Big).\label{npcond}
\eea
We split the field $\tF_{\mu\nu}$ into two parts
\bea
\tF_{\mu\nu}=B_{\mu\nu} + F_{\mu\nu}, 
\eea
where the background field $B_{\mu\nu}=\pro_\mu B_\nu-\pro_\nu B_\mu$ describes the magnetic vacuum gluon condensate, 
and 
$F_{\mu\nu}=\pro_\mu A_\nu-\pro_\nu A_\mu$ contains an Abelian potential which describes the primary gluon
 interacting  with the vacuum gluon condensate We will treat the potential $A_\mu$ as a wave function of a pure glueball
 formed as a system of the interacting primary gluon and corresponding vacuum gluon condensate.
We decompose the Lagrangian ${\cal L}^{\scriptsize\mbox{1-l}} $
around the vacuum condensate field $B_{\mu\nu}$ 
and obtain an effective Lagrangian for physical Abelian glueballs in the lowest quadratic approximation
\bea
{\cal L}_{\rm eff}^{(2)}[A]=-2 k_0 g^2 \dfrac{(B^{\mu\nu} F_{\mu\nu})^2}{B^2} \equiv
-\kappa (B^{\mu\nu} F_{\mu\nu})^2 , \label{LagrBf}
\eea
where we neglect a term corresponding to an absolute
value of the vacuum energy, and $\kappa$ is a free parameter.
The effective Lagrangian ${\cal L}_{\rm eff}^{(2)}[A]$ is strikingly different from the effective Lagrangians
obtained in quantum electrodynamics. Namely, the expression (\ref{LagrBf}) does not contain the 
classical kinetic term $-1/4 F_{\mu\nu}^2$ which is disappeared due to the  non-perturbative origin
of  the vacuum gluon condensate (\ref{npcond}) realizing the minimum of the effective potential.

Consider a case of the lightest magnetic glueball which is formed from the primary gluon in the presence of 
vacuum gluon condensate described by the vector harmonic $\vec A_{l=1,m=0}^{\mathfrak{m}}$,
 ($\omega=\nu_{11}$), which contains one non-zero magnetic potential $B_\varphi$
\bea 
B_\varphi(x,\theta,\tau)=N_{11}x j_1(\nu_{11} x)\sin^2 \theta \sin(\nu_{11}\tau),
\eea
where $N_{11}$ is a renormalization constant, and we introduce dimensionless variables
$x=r/\nu_{11}, \tau=t/\nu_{11}$ , $\nu_{11}=2.74\cdots$ is the first antinode of the radial function
$ r j_1(r)$.
The lightest magnetic glueball is described by the gauge potential 
$A_\varphi(x,\theta,\tau)$ which assumed to be time-coherent 
to the vacuum condensate field
\bea
A_\varphi(x,\theta,\tau)=a(x,\theta) \sin(\nu_{11} \tau +\phi_0),
\eea
with a constant phase shift $\phi_0$ . 
A time-averaged effective Lagrangian ${\cal L}_{\rm eff}^{(2)}[A]$ 
leads to the following Euler equation for the coordinate function $a(x,\theta)$ 
\begin{align}
&-\tM r^2 \sin\theta \Big(\tM r^2 b_{03}^2 +r^2 b_{13} \xi b_{03}'
\nn \\
&\hspace*{20mm}+\xi b_{03} (r^2 b_{13}'-2 r b_{13}-2 b_{23} ) \Big) a_\varphi \nn
\\
&+\cot\theta\csc \theta \Big (2 b_{23}^2 \big (\cos(2\theta)-
3\big )+2r^2  b_{23}b_{13}' \sin^2\theta \nn
\\
&\hspace*{20mm}-4 r b_{23} b_{13} \sin^2\theta (r b_{23}'-4 b_{23}) \Big )
 \pro_\theta a_\varphi
\nn 
\\
&+4 b_{23}^2\cos\theta \cot\theta \pro_\theta^2 a_\varphi 
\nn 
\\
&+2 r^2 b_{13}\sin\theta 
\big(r^2 b_{13}'-r b_{13}-b_{23} \big)\pro_\theta a_\varphi 
\nn
\\
&+4 r^2 b_{13} b_{23} \cos\theta \pro_r \pro_\theta a_\varphi
+r^4 b_{13}^2 \sin\theta \pro_r^2 a_\varphi=0,
\end{align}
%
where $\xi\equiv \cos(2 \phi_0)/(2+\cos(2 \phi_0))$, and  $b_{\mu\nu}(r)$ are coordinate parts of 
the field strength $B_{\mu\nu}$. The equation looks quite complicate. Surprisingly, for the ground state the equation 
is separable and admits 
a spherically symmetric solution $a_\varphi(r, \theta)\equiv f(r)$ which describes a glueball state 
with zero total angular momentum.
With this one results in an ordinary differential equation for the radial function $f$ (in dimensionless variables)
\begin{align}
x^2 b'^2 f'' -2b'(b+x b'-x^2 b'')f'& \nn \\
-\nu_{11}^2 \Big (\nu_{11}^2 x^2 b^2-\xi \nu_{11}^2 x^3 j_2(\nu_{11} x)  b' 2\Big ) f&=0, \label{eqf}
\end{align}
%
where $b(x)=\nu_{11} x j_1(\nu_{11}x)$,  $\xi\equiv\cos(2\phi_0)/(2+\cos(2\phi_0))$.
The coefficient functions in front of the first and second derivative terms in (\ref{eqf}) 
vanish at $x_0=1$ (or $r_0=\nu_{11})$. This implies localization of the solution 
in a finite interval $0\leq x \leq 1$, FIG. 8(a). 
Since the point $r_0=\nu_{11}$ represents a singularity it is suitable to apply
the ``shooting'' numeric method which allows to verify the type of singularity $r_0=\nu_{11}$. 
The regular structure of the solution has been checked 
in the small vicinity of the singularity $ r < \nu_{11} -1.0 \cdot 10^{-6}=2.743705$, and implies that
singularity belongs to removable type. This provides a smooth structure
of the energy density which has vanished first and second radial derivatives at $r_0$, FIG. 8(c),(d).
The solution has a removable singularity at $r=\nu_{11}$ (or $x_0=1$).
 To verify that solution is physical we 
check the properties of the energy density averaged over the time and polar angle
\begin{align}
\bar{\cal E}&=\dfrac{\pi \kappa}{4 x^2 } \Big ((2+\cos(2 \phi_0)) b^2 f^2 +2 \cos(2 \phi_0)  b b' f f' \nn \\
&\hspace*{13mm} +(2+\cos(2 \phi_0)) b'^2 f'^2 \Big ).
\end{align}
\begin{figure}[h!]
\centering
{\includegraphics[width=70mm,height=52mm]{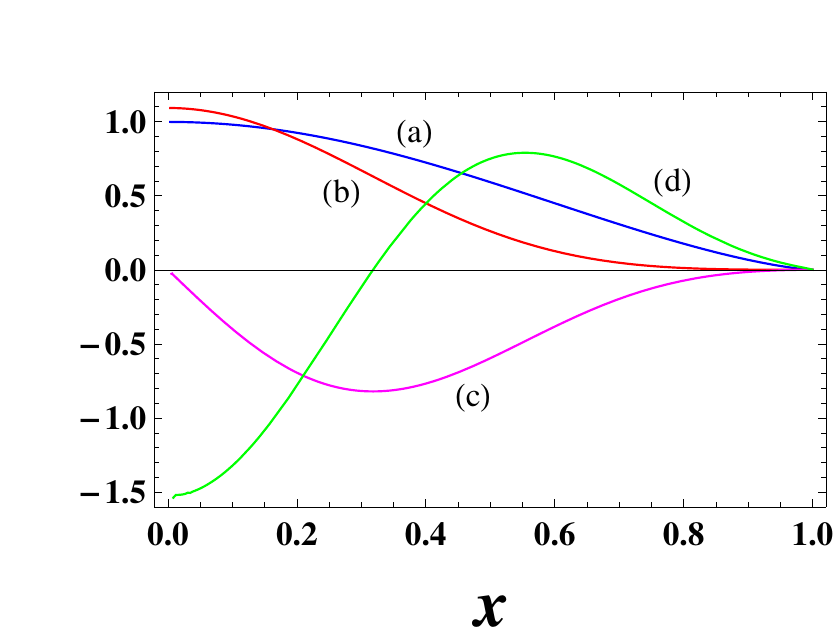}}
\caption[fig8]{(a) Solution $f(x)$ with a unit amplitude(in blue); (b) a corresponding radial energy density 
$\bar{\cal E}/4 \pi \kappa$(in red); 
(c) the first derivative of the energy density(in violet); (d) the second derivative of the energy density (in green); $\phi_0=\pm \pi/2$.      
}\label{Fig8}
\end{figure} 
The solution has a minimal energy at the phase shift $\phi_0=\pi/2$, and only at this value
the averaged over time effective Lagrangian vanishes completely, like in a case of free Lagrangian for photon plane waves.
 So the obtained solution describes a stable ground state for a scalar glueball.
Quantum numbers of the glueball can be defined in the same way as for two free photons system 
and lead to two lightest glueballs $0^{++}, 0^{-+}$ \cite{kochelev2009,ochs2013}.

Note that result for the effective Lagrangian (\ref{LagrBf}) is model independent,
and it can be obtained from a class of Lagrangian functions (like in Ginsburg-Landau model) which 
admit series expansion around a non-trivial vacuum. 
Qualitative estimates of the lightest scalar glueball spectrum
can be performed in a model independent way assuming that
vacuum gluon condensate is a universal order parameter for glueballs with different quantum numbers.
The knowledge of explicit solutions for the vector potential (\ref{vecharmonics})
allows to find analytical expressions for the radial density of the vacuum gluon condensate functions performing 
averaging over the time and polar angle.
Averaged over the time and polar angle vacuum gluon condensate functions
  $\alpha_s\langle (F_{\mu\nu})^2 \rangle$ corresponding to magnetic modes 
  $\vec A_{11}^{\mathfrak{m}}(\nu_{11} x)$ and  $\vec A_{ 11}^{\mathfrak{m}}(\mu_{11} x)$
    are depicted in FIG. 8 ($\alpha_s=0.5$).
 \begin{figure}[h!]
{\includegraphics[width=68mm,height=42mm]{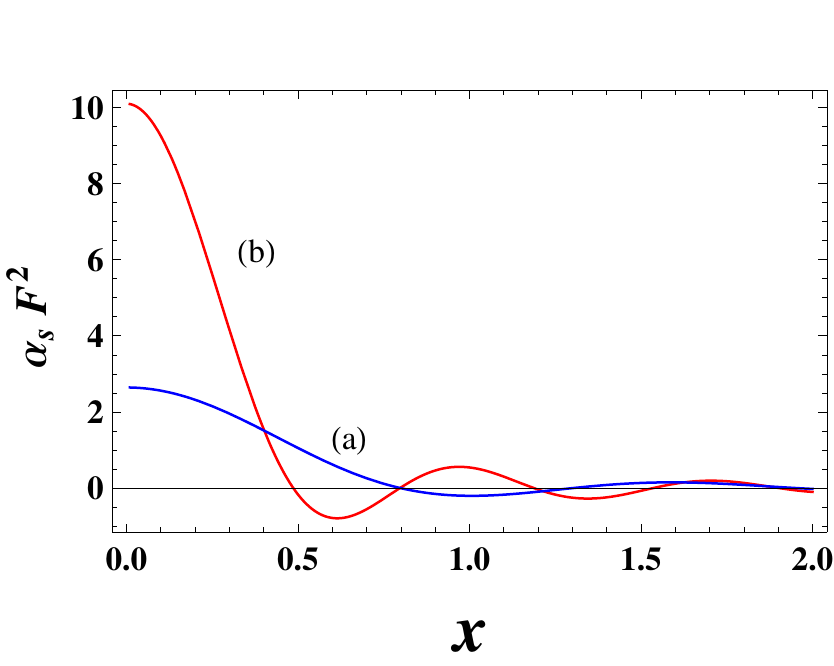}}
\caption[fig9]{Radial densities of the magnetic vacuum gluon condensates $\alpha_s\langle \overline{F^2} \rangle$ 
corresponding to modes $\nu_{11}=2.74\cdots$, (a), and $\mu_{11}=4.49\cdots$, (b); $(n=l=1, m=0)$.}\label{Fig9}
\end{figure}
The oscillating behavior of the vacuum gluon condensate density was obtained before within 
the instanton approach to QCD \cite{dorokhov1997}.

 Integrating the radial density over the interval $(0\leq x \leq 1)$ one can fit 
a value of the obtained vacuum gluon condensate parameter to the known value 
$\alpha_s\langle (F_{\mu\nu}^a)^2 \rangle=(540{\rm [MeV]})^4$,  
and obtain an explicit dependence of the glueball size on quantum number $\tM_{nl}$
\begin{align}
a_{nl}[{\rm fm}]&=\dfrac{197}{v_0} f_{c}^{1/4}(\tM) \approx \dfrac{107 \alpha_s^{1/4}}{v_0}\sqrt{\tM_{nl}},  \label{aM} \\
 f_{c}(\tM)&=\dfrac{N_{nl}^2}{12\pi \tM^2}\big((3-4 \tM^2+2\tM^4)\cos(2 \tM)-3\nn \\
&-2 \tM^2+2 \tM (3-\tM^2) \sin(2 \tM)+4 \tM^5 {\rm si} (2 \tM)\big),
\nn
\end{align}
where $v_0=540[1/{\rm fm}]$,  $N_{nl}$ is the normalization factor of the vector harmonic, 
and ${\rm si} (2\tM)$ is the sine integral function.
With this one can find the energy spectrum of light scalar glueballs $J^{PC}=0^{+ +}$
\bea
E_{nl}[{\rm MeV}]=\tk v_0 f_c^{-1/4} \tM_{nl}\approx
 \Big(\dfrac{80}{7 \alpha_s}\Big )^{1/4} \tk v_0 \sqrt {\tM_{n1}},  \label{EnM}
\eea
where $\tk$ is a free model parameter which can be fixed by fitting the energy value of the lightest glueball.
For $\tk=1.01$ the lightest glueball $J^{PC}=0^{++}$ has 
energy $E_{\nu_{11}}=1440[{\rm MeV}]$. The energy spectrum (\ref{EnM})
agrees with the Regge theory of hadrons. 

\section{Discussion}
      We have demonstrated that Weyl symmetric solutions provide color singlet primary quantum states
      for gluons and quarks. Instead of eight free color gluons defined in the framework of a perturbative QCD one has 
      an infinite number of primary gluons of magnetic or electric type 
      with quantum numbers $(l,m, k)$ which are localized in finite space domains constrained by nodes/antinodes $\tM_{nl}$.
       Physical observables, hadrons, are formed as systems of  interacting primary gluons and quarks with corresponding 
       generated vacuum gluon and quark condensates. Certainly, this implies that one has to construct an improved
       quark model of hadrons  which might be successful in resolving another persistent problem of proton mass and spin.
       This problem and other related issues will be considered elsewhere.

\acknowledgments

Authors thank A.B. Voitkiv, A. Silenko, J. Evslin,  S.-P. Kim, A. Kotikov, A. Pimikov and Ed. Tsoi for 
valuable discussions. This  work is supported by Chinese Academy of Sciences  (PIFI Grant No. 2019VMA0035),
National Natural Science Foundation of China (Grant No. 11575254),
and by Japan Society for Promotion of Science (Grant No. L19559).
\\

{ \bf Appendix:  Singlet structure of Weyl symmetric solutions: a simple model}
\\

We demonstrate a source of singlet structure of Weyl symmetric solutions by solving a reduced system 
of equations for two long distance propagating modes
$K_2,K_4$ (\ref{eq2}), (\ref{eq3}). In the asymptotic region fields $K_{0,1}$ vanish and 
     one has in the leading order the following factorized structure of solutions for $K_{2,4}$
     \be
\begin{array}{rcl}
     K_2(r,\theta,t)\!\!&=&\!\!f_2(M r) T_2(\theta) \cos(M t), \\ [2\jot]
      K_4(r,\theta,t)\!\!&=&\!\!f_4(M r) T_4(\theta) \cos(M t).
\end{array}
      \ee
      We set for simplicity $M=1$, and consider the lowest energy solution
      with quantum number  $J=l=1$ and lowest polar modes $T_2(\theta)=1,~T_4(\theta)=\sin^2 \theta.$
    With this, the equations for $K_{2,4}$ are simplified as follows
    \be
\begin{array}{rcl}
   f''_2+f_2-\dfrac{9}{4r^2} f_2 f_4^2\sin^2 \theta\!\!&=&\!\!0,  \\ [3\jot]
   f''_4+f_4-\dfrac{2}{r^2} f_4-\dfrac{3}{2 r^2}f_4 f_2^2\cos^2\theta\!\!&=&\!\!0.
\end{array}
    \ee
   To obtain qualitative estimate we perform averaging over polar angle which leads to
   a simple system of ordinary differential equations  
           \be
\begin{array}{rcl}
      \tf''_2+\tf_2-\dfrac{1}{r^2} \tf_2 \tf_4^2\!\!&=&\!\!0,  \\ [3\jot]
    \tf''_4+\tf_4-\dfrac{2}{r^2}f_4-\dfrac{1}{r^2} \tf_4 \tf_2^2\!\!&=&\!\!0,
\end{array}
    \ee
    where $\tf_2=\sqrt 3/2 f_2$, $\tf_4=3/2 \sqrt 2 f_4$.
\begin{figure}[h!]
\centering
\subfigure[~]{\includegraphics[width=66mm,height=41mm]{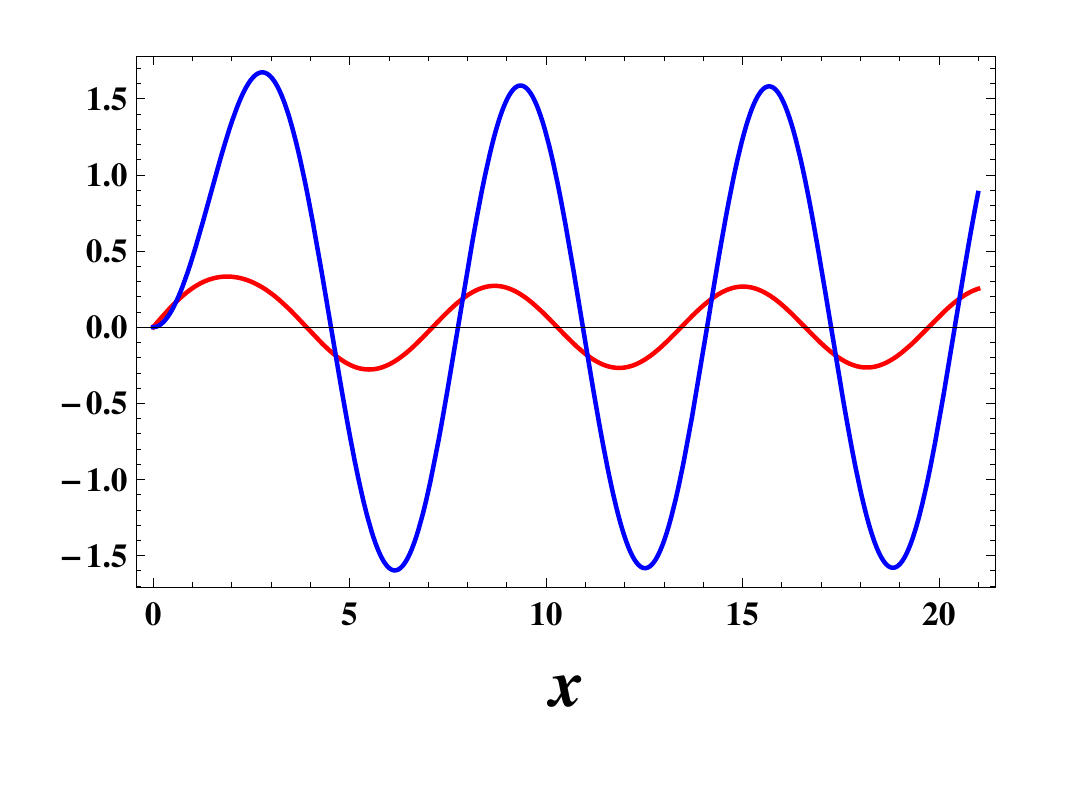}}
\subfigure[~]{\includegraphics[width=66mm,height=42mm]{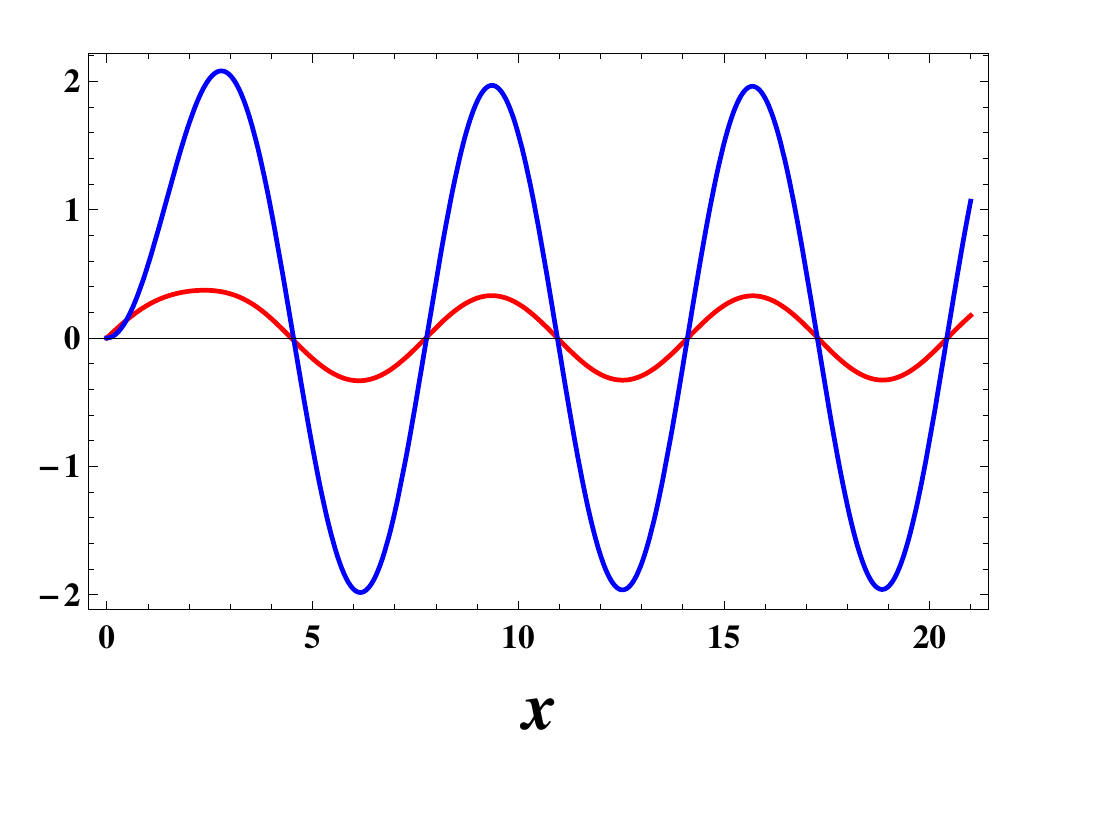}}
\caption[fig10]{Solution for $\tf_2$ (in red), $\tf_4$ (in blue): (a) $\tf_2,\tf_4$ do not have common
zeros; (b) $\tf_{2,4}$ have the same zeros and extremums.}\label{Fig10}
\end{figure}
    All possible solutions to that system of equations can be easily obtained 
    by applying  a ``shooting'' numeric method by setting initial values and first derivatives 
    of functions $\tf_{2,4}$ at the origin $r=0$. Local solution near the origin $r=0$ implies
    that $\tf_{2,4}(0)=0$, so a general solution is defined by two integration constants corresponding 
    to normal derivatives $\tf'_{2,4}(0)=c_{2,4}$, and the space of solutions with given quantum numbers
    is two-dimensional in agreement with a simple counting degrees of freedom in the presence of local $U(1)$ symmetry.
     An example of a general solution is presented in FIG. 4(a). Nodes and extremums of a general solution profile
    functions $\tf_2,\tf_4$ do not coincide, so that such solutions do not possess a conserved 
    energy in any finite region. Moreover, such solutions are not classically stable under small
    fluctuations and represent saddle points. Stable stationary solutions with a conserved localized energy are 
    selected by a constraint that fields $\tf_{2,4}$ must have the same, at least one node or antinode., FIG. 4(b). 
    Location of nodes/antinodes of fields $\tf_{2,4}$ is determined by two integration constants $c_{2,4}$. Two fields
    $\tf_{2,4}$ admit common nodes for special values of $c_{2,4}$ which imply correlation of amplitudes of $\tf_{2,4}$.
    Therefore, a space of solutions with a given quantum numbers $M,l,m,k$ and conserved energy in a finite space region 
     is one-dimensional and the norm of solution is determined by one normalization constant which can be assign
     to  the amplitude of the Abelian field $K_4$.This provides color singlet quantum states after quantization.
Due to classical electric-magnetic duality one has similar results 
for electric type solutions. 

\end{document}